\documentclass[useAMS,usenatbib]{mn2e}
\usepackage{epsfig}
\usepackage{color}
\usepackage{subfigure}
\usepackage{amsmath}
\usepackage{amssymb}

\def\lsim{\mathrel{\rlap{\lower3.5pt\hbox{\hskip0.5pt$\sim$}}
    \raise0.5pt\hbox{$<$}}}
\def\gsim{~\rlap{$>$}{\lower 1.0ex\hbox{$\sim$}}}

\newcommand{\goodgap}{\hspace{\subfigtopskip} \hspace{\subfigbottomskip}}

\title[Peaks statistics and $f(R)$ theories]{Weak lensing peak count as a probe of $f(R)$ theories}
\author[V.F. Cardone et al.]{V.F. Cardone$^{1}$\thanks{Corresponding author\,: {\tt winnyenodrac@gmail.com}}, S. Camera$^2$, R. Mainini$^3$, A. Romano$^1$, A. Diaferio$^{4,5}$, R. Maoli$^6$, \and R. Scaramella$^1$ \\
$^1$I.N.A.F.\,-\,Osservatorio Astronomico di Roma, via Frascati 33, 00040 - Monte Porzio Catone (Roma), Italy \\
$^2$CENTRA, Instituto Superior T\'ecnico, Universidade T\'ecnica de Lisboa, Av. Rovisco Pais 1, 1049\,-\,001 Lisboa, Portugal \\
$^3$Dipartimento di Fisica "G. Occhialini", Universit\`{a} di Milano\,-\,Bicocca, Piazza della Scienza 3, 20126 - Milano, Italy \\
$^4$Dipartimento di Fisica, Universit\`a degli Studi di Torino, Via Giuria 1, 10125 - Torino, Italy \\
$^5$Istituto Nazionale di Fisica Nucleare, Sezione di Torino, Via Giuria 1, 10125 - Torino, Italy \\
$^6$Dipartimento di Fisica, Universit\`{a} di Roma "La Sapienza", Piazzale Aldo Moro, 00185 - Roma, Italy \\}

\date{Accepted xxx, Received yyy, in original form zzz}

\begin{document}

\maketitle

\begin{abstract}

Weak gravitational lensing by galaxy clusters on faint higher redshift galaxies has been traditionally used to study the cluster mass distribution and as a tool to identify clusters as peaks in the shear maps. However, it becomes soon clear that peaks statistics can also be used as a way to constrain the underlying cosmological model due to its dependence on both the cosmic expansion rate and the growth rate of structures. This feature makes peak statistics particularly interesting from the point of view of discriminating between General Relativity and modified gravity. Here we consider a general class of $f(R)$ theories and compute the observable mass function based on the aperture mass statistics. We complement our theoretical analysis with a Fisher matrix forecast of the constraints that an Euclid\,-\,like survey can impose on the $f(R)$ model parameters. We show that peak statistics alone can in principle discriminate between General Relativity and $f(R)$ models and strongly constrain the $f(R)$ parameters that are sensitive to the non-linear growth of structure. However, further analysis is needed in order to include possible selection function in the peaks redshift determination.

\end{abstract}

\begin{keywords}
cosmology\,: theory -- gravitational lensing -- clusters\,: general
\end{keywords}

\section{Introduction}

The gravitational field of galaxy clusters distorts the images of background faint galaxies possibly leading to the formation of spectacular giant arcs (as first observed by \citealt{S87}). As pointed out by \cite{W85} in a pioneering work, the most likely effect is, however, a variation in the ellipticity distribution of the background galaxies which can then be used to reconstruct the mass distribution of the lensing cluster (see, e.g., \citealt{KS93,KSB95}). Such a mass reconstruction method has now become quite popular \citep{Clowe98,Mario2008,Anna10,Zhuoyi11} and is actually the most efficient one for intermediate and high redshift clusters where dynamical methods fail because of the difficulties in measuring galaxy redshits.

The search for coherent image alignments can also be used as a way to find dark mass concentrations thus offering the opportunity of assembling a mass selected catalog of haloes. In particular, the aperture mass statistics \citep{S96} has emerged as a valuable way to find clusters \citep{H05,GS07,Sch07} and measuring their mass function \citep{D06}. On the other hand, although not originally conceived as a cosmological tool, the aperture mass statistics allows to severely constrain cosmological models. Indeed, the number counts of peaks in the weak lensing maps is determined by both the cosmic expansion rate and the growth rate of structures (entering through its effect on the theoretical mass function). Peaks statistics has therefore emerged as a promising tool to discriminate among different dark energy models \citep{BPB02,M09,DH10,M10,K10} and constrains primordial non\,-\,Gaussianity \citep{MFM11}.

Actually, dark energy is not the only way to fit the wide amount of data indicating the present day universe as spatially flat and undergoing accelerated expansion \citep{W12}. Indeed, rather than representing the evidence of a missing source in the cosmic energy budget, cosmic speed up may also be read as the first signal of breakdown of our understanding of the laws of gravity. Modified gravity theories has therefore recently attracted a lot of interest and have been shown to be able to fit the data with the same accuracy as most dark energy models. In this framework, $f(R)$ theories represent one of the most natural generalization of General Relativity, the basic idea being to replace the scalar curvature $R$ in the gravity Lagrangian with a generic function $f(R)$. Dating back to Eddington, this idea has been first reconsidered in the `80s thanks to the possibility of recovering inflation without any scalar field and has then found a renewed interest after the discovery of the acceleration for its ability to achieve cosmic speed up in a matter only universe (see, e.g., \citealt{dFT10,CdL11} and refs. therein). It is actually possible to show that, for any dark energy model, it is possible to work out an $f(R)$ counterpart providing the same background expansion, i.e. the same $H(z)$, as the given one \citep{CCT05,MV06}. On the contrary, these equivalent models can be discriminated because of the different growth of perturbations driven by a scale and redshift dependent effective gravitational constant and the non vanishing difference between the two Bardeen potentials.

As already reminded above, the number counts of weak lensing peaks is indeed sensible to both the expansion rate and the growth of structures so that it comes out as an ideal tool to discriminate among dark energy and $f(R)$ models sharing the same $H(z)$ expression. Here, we will therefore investigate whether this is indeed the case for a popular $f(R)$ model which has already been shown to excellently fit a wide astrophysical dataset. Such a preliminary study can highlight the power of peaks statistics in discriminating among dark energy and modified gravity, but it is not immediately related to what one could actually infer from observations. We will therefore rely on the Fisher matrix analysis to forecast the accuracy a realistic survey can achieve on the $f(R)$ model parameters thus filling (although in a simplified way) the gap between theory and observations.

The plan of the paper is as follows. In Sect.\,2 we review the basic of $f(R)$ theories and how one can compute the mass function taking care of the peculiarities of this modified gravity model. Sect.\,3 is then devoted to the mass aperture statistics discussing all the steps needed to compute both the signal and the noise in weak lensing maps. The expected number of counts of peaks in $f(R)$ theories and how this depends on the model parameters are presented in Sect.\,4, while the Fisher matrix forecasts are given in Sect.\,5. Sect.\,6 summarizes the main results and our conclusions.

\section{Fourth order gravity}

Being a straightforward generalization of Einstein GR, fourth order gravity theories have been investigated almost as soon as the original Einstein theory appeared. It is, therefore, not surprising that the corresponding field equations and the resulting cosmology have been so widely discussed in the literature. We here first briefly review the basics of $f(R)$ theories and then explain the method used to evaluate the theoretical mass function.

\subsection{$f(R)$ cosmology}

In the framework of the metric approach, the field equations are obtained by varying the gravity action

\begin{equation}
S = \int{d^4x \sqrt{-g} \left [ \frac{f(R)}{16 \pi G} + {\cal{L}}_M \right ]}
\label{eq: fogaction}
\end{equation}
with respect to the metric components. We obtain

\begin{equation}
f^{\prime} R_{\mu \nu} - \nabla_{\mu \nu} f^{\prime} + \left ( \square f^{\prime} - \frac{1}{2} f \right ) g_{\mu \nu} = 8 \pi G T_{\mu \nu} \ ,
\label{eq: fogfield}
\end{equation}
where $R$ is the scalar curvature, ${\cal{L}}_M$ is the standard matter Lagrangian with $T_{\mu \nu}$ the matter stress\,-\,energy tensor, and the prime (the dot) denotes derivative with respect to $R$ (time $t$). For $f(R) = R - 2 \Lambda$, one obtains the usual Einstein equations with a cosmological constant $\Lambda$, while, in the general case, a further scalar degree of freedom is introduced.

In a spatially flat homogenous and isotropic universe, some convenient algebra allows to rearrange the field equations in such a way that a single equation for the Hubble parameter $H = \dot{a}/a$ is obtained. Assuming dust as gravity source and introducing $E = H(z)/H_0$, it is then only a matter of algebra to get\,:

\begin{equation}
E^2(z) = \frac{\Omega_M \left [ (1 + z)^3 + (\xi f^{\prime} - m^2 f)/6 \right ]} {f^{\prime} - m^2 (1 + z) (d\xi/dz)} \ ,
\label{eq: hubevsxi}
\end{equation}

\begin{displaymath}
m^2 f^{\prime \prime} \frac{d^2 \xi}{dz^2} + m^4 f^{\prime \prime \prime} \left ( \frac{d\xi}{dz} \right )^2 - \left [2 - \frac{d\ln{E(z)}}{d\ln{(1 + z)}} \right ] \frac{m^2 f^{\prime \prime}}{1 + z} \frac{d\xi}{dz}
\end{displaymath}
\begin{equation}
= \frac{\Omega_M}{E^2(z)} \left [ (1 + z) - \frac{2 m^2 f - \xi f^{\prime}}{3 (1 + z)^2} \right ] \ ,
\label{eq: xizeq}
\end{equation}
where $z = 1/a - 1$ is the redshift, $\xi = R/m^2$ and

\begin{equation}
m^2 = \frac{(8 \pi G)^2 \rho_M(z = 0)}{3} \simeq (8315 \ {\rm Mpc})^{-2} \ \left ( \frac{\Omega_M h^2}{0.13} \right )
\label{eq: defm}
\end{equation}
is a convenient curvature scale which depends on the present day values of the matter density parameter $\Omega_M$ and the Hubble constant $h = H_0/(100 \ {\rm km/s/Mpc})$.

By inserting Eq.(\ref{eq: hubevsxi}) into Eq.(\ref{eq: xizeq}), we get a single second order nonlinear differential equation for $\xi(z)$ that can be solved numerically provided $f(R)$ and the initial conditions are given. These latter can be conveniently expressed in terms of the present day values of the deceleration ($q = -H^{-2} \ddot{a}/a$) and jerk ($j = H^{-3} \dddot{a}/a$) parameters. To this end, we first remember that, in a FRW spatially flat universe, the curvature scalar $R$ reads\,:

\begin{displaymath}
R = 6 (\dot{H} + 2 H^2)
\label{eq: defR}
\end{displaymath}
so that we get for the present-day values \citep{CCS08}\,:

\begin{displaymath}
R_0 = 6 H_0^2 (1 - q_0) \ \ , \ \ \dot{R}_0 = 6 H_0^3 (j_0 - q_0 - 2)  \ \ .
\end{displaymath}
It is then only a matter of algebra to get the initial conditions for $\xi$ as

\begin{equation}
\left \{
\begin{array}{l}
\displaystyle{\xi(z = 0) = (6/\Omega_M) (1 - q_0)} \\
~ \\
\displaystyle{d\xi/dz(z = 0) = (6/\Omega_M) (j_0 - q_0 - 2)} \\
\end{array}
\right . \ .
\label{eq: incond}
\end{equation}
Because of the definition of $\xi$, Eq.(\ref{eq: xizeq}) is a single fourth\,-\,order nonlinear differential equation for the scale factor $a(t)$ so that we need to know the values of the derivatives up to the third order to determine the evolution of $a(t)$ thus explaining why the jerk parameter also enters as a model parameter.

A key role in fourth order theories is obviously played by the functional expression adopted for $f(R)$. We choose here the Hu \& Sawicki (2007, hereafter HS) model setting\,:

\begin{equation}
f(R) = R - m^2 \frac{c_1 (R/m^2)^n}{1 + c_2 (R/m^2)^n}
\label{eq: frhs}
\end{equation}
where $m^2$ is given by (\ref{eq: defm}), and $(n, c_1, c_2)$ are positive dimensionless constants. Note that, since

\begin{displaymath}
\lim_{m^2/R \rightarrow 0}{f(R)} \simeq R - \frac{c_1}{c_2} m^2 + \frac{c_1}{c_2^2} m^2 \left ( \frac{m^2}{R} \right )^n \ ,
\end{displaymath}
we recover an effective $\Lambda$ term in high curvature $(m^2/R \rightarrow 0)$ environments. In particular, in the limit $R >> m^2$, the expansion rate $H$, in the early universe, will be the same as in $\Lambda$CDM with an effective matter density parameter $\Omega_{M,eff} = 6 c_2/(c_1 + 6 c_2)$ that guarantees that the nucleosynthesis constraints are satisfied.

The HS model is determined by the three parameters $(n, c_1, c_2)$, but it is actually convenient to replace $(c_1, c_2)$ with $(q_0, j_0)$. To this end, we follow the method detailed in \cite{CCD11} and introduce the further parameter $\varepsilon = \log{(f_{R0} - 1)} = \log{[f^{\prime}(z = 0)- 1]}$ which quantifies the present day deviation of $f(R)$ from GR. Note that, since we expect $f_{R0} - 1 << 1$, $\varepsilon$ will typically be negative and large.

Inserting the HS $f(R)$ into Eqs.(\ref{eq: hubevsxi}) and (\ref{eq: xizeq}) and setting the model parameters, one can solve the fourth order nonlinear differential equation for the scale factor $a(t)$ and then work out the dimensionless Hubble rate $E(z)$. In order to speed up the computation, however, we use here the following approximated solution \citep{CCD11}\,:

\begin{equation}
E(z) = \left \{
\begin{array}{ll}
{\cal{E}}(z) E_{CPL}(z) + [1 - {\cal{E}}(z)] E_{\Lambda}(z) & z \le z_{\Lambda} \\
~ & ~ \\
E_{\Lambda}(z) & z \ge z_{\Lambda} \\
\end{array}
\right .
\label{eq: hubapprox}
\end{equation}
where

\begin{equation}
E^2_{CPL} = \Omega_M (1 + z)^3 + (1 - \Omega_M) (1 + z)^{3(1 + w_0 + w_a)} {\rm e}^{- \frac{3 w_a z}{1 + z}}
\label{eq: ecpl}
\end{equation}
is the Hubble rate for the phenomenological CPL \citep{CP01,L03} parametrization of the effective dark energy fluid equation of state, namely $w(z) = w_0 + w_a (1 - a)$, while $E_{\Lambda}(z) = E_{CPL}(z, w_0 = -1, w_a = 0)$. Finally

\begin{equation}
{\cal{E}}(z) = \sum_{i = 1}^{3}{e_i (z - z_{\Lambda})^{i}}
\label{eq: defepsfun}
\end{equation}
is an interpolating function with $e_i$ and $z_\Lambda$ depending on $(\Omega_M, q_0, j_0, n, \varepsilon)$. This approximating function excellently reproduces the numerical solution whatever the model parameters $(\Omega_M, q_0, j_0, n, \varepsilon)$ are with a rms error which is far lower than $0.1\%$ over the redshift range $(0, 1000)$. A subtle remark is in order here concerning the value of $\Omega_M$. Indeed, while for $z \le z_{\Lambda}$, we use the actual matter density parameter, the effective one must be used for $z \ge z_{\Lambda}$. Therefore, a discontinuity in $z_{\Lambda}$ is formally present in our approximation. Actually, it is easy to show that, for all reasonable model parameters, $\Omega_M$ and $\Omega_{M,eff}$ are almost perfectly equal so that the discontinuity can not be detected at all and $E(z)$ is, to all extents, a continuous function.

\subsection{Mass function for $f(R)$ gravity}

The extra scalar degree of freedom introduced by the modified gravity Lagrangian mediates an enhanced gravitational force on scales smaller than its Compton wavelength. In order to hide this boost from local tests of gravity, viable $f(R)$ models resort to the chameleon mechanism \citep{MB04,KW04} which makes the field Compton wavelength depending on the environment gravitational potential. As a consequence, for viable $f(R)$ models, the gravitational potential reduces to the Newtonian one on Solar System scales, while nonlinearities in the field equations appear for galaxy scale systems. From the point of view of the mass function, the enhanced gravitational force has a particularly strong impact on the abundance on intermediate mass haloes. Indeed, on one hand, the extra force increases the merging rate of low mass haloes into intermediate ones, while the chameleon shuts down the merging of these latter into highest mass ones.

\begin{figure*}
\centering
\subfigure{\includegraphics[width=7.5cm]{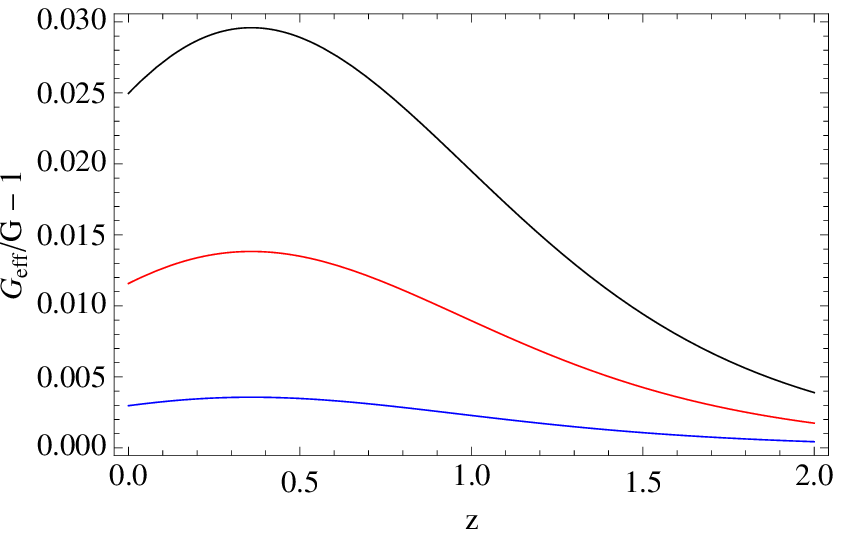}} \goodgap
\subfigure{\includegraphics[width=7.5cm]{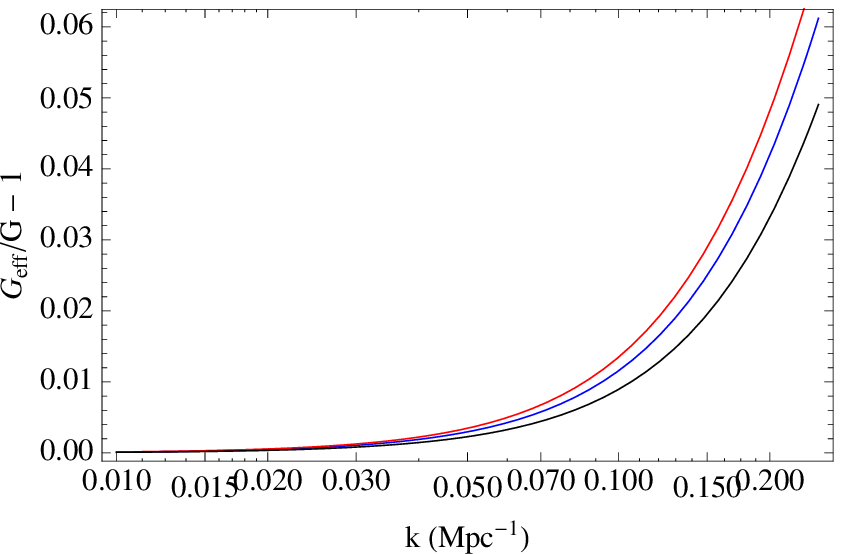}} \goodgap
\caption{Effective gravitational constant for the fiducial HS model as a function of $z$ (left panel) for $k = 0.05$ (blue), 0.10 (red), 0.15 (black) and $k$ (right panel) for $z = 0$ (blue), 0.5 (red), 1.0 (black).}
\label{fig: geff}
\end{figure*}

An efficient method to take the chameleon effect into account and work out a mass function in agreement with simulations has been recently proposed by \cite{LH11} based on a simple interpolation scheme. Following their approach, we model the mass function as\,:

\begin{equation}
{\cal{N}}(\ln{M_{vir}}) = \frac{\rho_M(z = 0)}{M_{vir}}  \ \frac{d\ln{\nu}}{d\ln{M_{vir}}}  \ \nu \varphi(\nu) \  ,
\label{eq: mf}
\end{equation}
where $\varphi(\nu)$ is the number of halos with the ratio between density contrast and variance of the perturbations equal to $\nu = \delta_c/\sigma(M_{vir})$, $\delta_c$ is the critical overdensity for collapse and $\sigma$ the variance of the perturbations on the scale $R$ corresponding to the mass $M_{vir}$, given by

\begin{equation}
\sigma^2[R(M_{vir})] = \frac{1}{(2 \pi)^3} \int{P_{\delta}(k) |W(kR)|^2 d^3k} \ .
\label{eq: sigmavardef}
\end{equation}
Here, $W(kR)$ is the Fourier transform of the spherical top hat function, while the density power spectrum $P_{\delta}(k)$ may be computed as\,:

\begin{equation}
P_{\delta}(k, z) = {\cal{A}} k^{n_{PS}} T^2(k) D^2(k, z)
\label{eq: powspec}
\end{equation}
with $T(k)$ the transfer function, $D(k, z) = \delta(k, z)/\delta(k, 0)$ the growth factor normalized to 1 at present day and ${\cal{A}}$ a normalization constant conveniently expressed as function of $\sigma_8$, the present day variance on the scale $R = 8 h^{-1} \ {\rm Mpc}$. Since $f(R)$ models reduce to GR in the prerecombination epoch, we can use for $T(k)$ the analytical approximation in \cite{EH98} although this latter has been obtained assuming GR validity. On the contrary, the growth factor for $f(R)$ models gains a scale dependence which is not present for GR dark energy models. Indeed, the evolution of the density perturbations, $\delta = \delta \rho_M/\rho_M$, in fourth order gravity models is ruled by

\begin{equation}
\ddot{\delta} + 2 H \dot{\delta} - 4 \pi {\cal{G}}_{eff}(a, k) \rho_M \delta = 0
\label{eq: growtheq}
\end{equation}
with $k$ the wavenumber and

\begin{figure*}
\centering
\subfigure{\includegraphics[width=7.5cm]{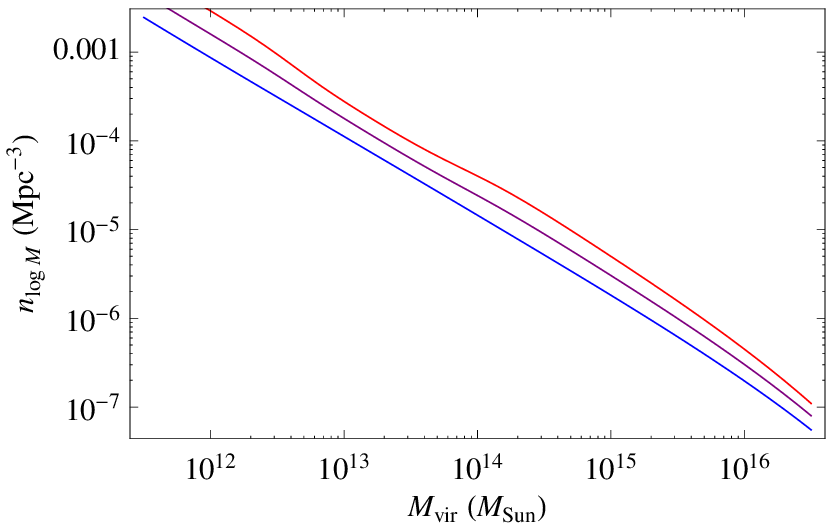}} \goodgap
\subfigure{\includegraphics[width=7.5cm]{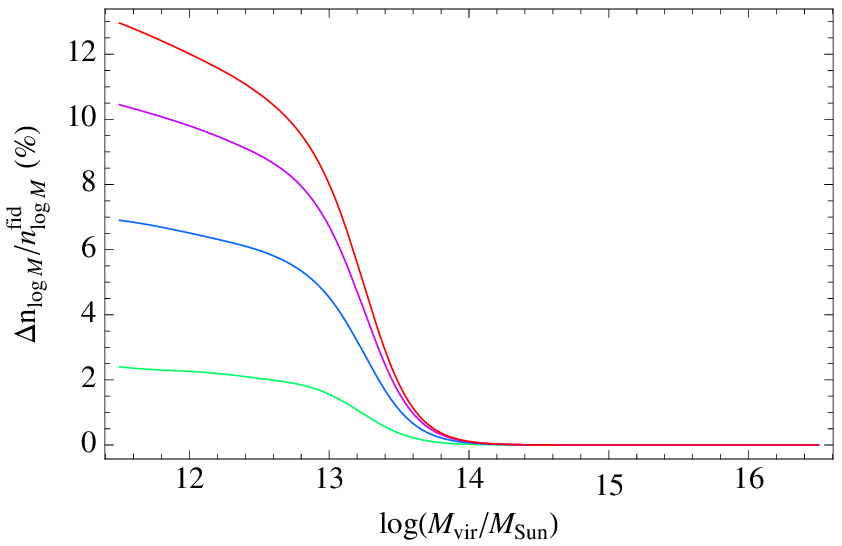}} \goodgap
\caption{{\it Left.} Mass function (MF) for the fiducial HS model for $z = 0$ (blue), $z = 0.5$ (purple), $z = 1.0$ (red) shifting by 0.25 (0.50) dex the MF at $z = 0.5$ ($z = 1.0$) for a better visualization. {\it Right.} Pecentage deviation, $\Delta {\cal{N}}(\log{M_{vir}})/{\cal{N}}(\log{M_{vir}}) = [{\cal{N}}_{HS}(\log{M_{vir}}) - {\cal{N}}_{CPL}(\log{M_{vir}})]/{\cal{N}}_{HS}(\log{M_{vir}})$,  as a function of the halo mass for $z$ from 0.2 to 1.4 in steps of 0.4 (green, blue, purple and red lines).}
\label{fig: mffid}
\end{figure*}

\begin{equation}
{\cal{G}}_{eff}(a, k) = \frac{G}{f^{\prime}(R)} \frac{1 + 4 (k^2/a^2) [f^{\prime \prime}(R)/f^{\prime}(R)]}{1 + 3 (k^2/a^2) [f^{\prime \prime}(R)/f^{\prime}(R)]}
\label{eq: geffdef}
\end{equation}
the scale dependent effective gravitational constant \citep{T07}. Note that, for $f(R) = R - 2 \Lambda$, ${\cal{G}}_{eff}$ reduces to $G$ and the growth factor is no more scale dependent. As shown in Fig.\,\ref{fig: geff}, setting the model parameters to the fiducial values, $G_{eff}$ only slightly deviates from the Newton constant. Nevertheless, this deviation is both redsfhift and scale dependent so that it has an impact on the final matter power spectrum.

In order to compute the mass function through Eq.(\ref{eq: mf}), one has to choose an expression for $\nu \varphi(\nu)$. We again follow \cite{LH11} and adopt the \cite{ST99} function setting

\begin{equation}
\nu \varphi(\nu) = A \sqrt{\frac{2 a \nu^2}{\pi}} [1 + (a \nu^2)^{-p}] \exp{(-a \nu^2/2)}
\label{eq: stmf}
\end{equation}
with $(A, a, p) = (0.322, 0.75, 0.3)$. It is, finally, worth stressing that both $\delta_c$ and $\Delta_{vir}$ (which enters the conversion from $M_{vir}$ to $R_{vir}$) actually depends on the matter density parameter $\Omega_M$ and the physics ruling the collapse and virialization of perturbations. Previous attempts to fit the numerically derived mass function searched for a $f(R)$ derivation of $(\delta_c, \Delta_{vir})$ and then use the mass variance $\sigma(M_{vir})$ evaluated from the $f(R)$ linear power spectrum. On the contrary, here we follow the approach of \cite{LH11} holding $(\delta_c, \Delta_{vir})$ fixed to the values predicted for the GR model having the same background expansion as the $f(R)$ one. We then use for the mass variance an interpolation between the $f(R)$ and GR ones, i.e. $\sigma(M_{vir})$ is estimated as\,:

\begin{equation}
\sigma(M_{vir}) = \frac{\sigma_{FoG}(M_{vir}) + (M_{vir}/M_{th})^{\alpha} \sigma_{GR}(M_{vir})}{1 + (M/M_{th})^{\alpha}}
\label{eq: sigmalh}
\end{equation}
with $\sigma_{FoG}$ and $\sigma_{GR}$ computed from the $f(R)$ and GR linear power spectra and $(\alpha, M_{th})$ two interpolation parameters given by

\begin{equation}
\left \{
\begin{array}{l}
\displaystyle{\alpha = 2.448} \\
~ \\
\displaystyle{M_{th} = 1.345 \times 10^{13} \ \left ( \frac{|f_{R0}| - 1}{10^{-6}} \right )^{3/2} \ h^{-1} \ {\rm M_{\odot}}} \\
\end{array}
\right . \ .
\end{equation}
Finally, we follow \cite{BN98} to get $(\delta_c, \Delta_{vir})$ as a function of $\Omega_M$ and the dimensionless Hubble rate $E(z)$.

\begin{figure*}
\centering
\subfigure{\includegraphics[width=7.5cm]{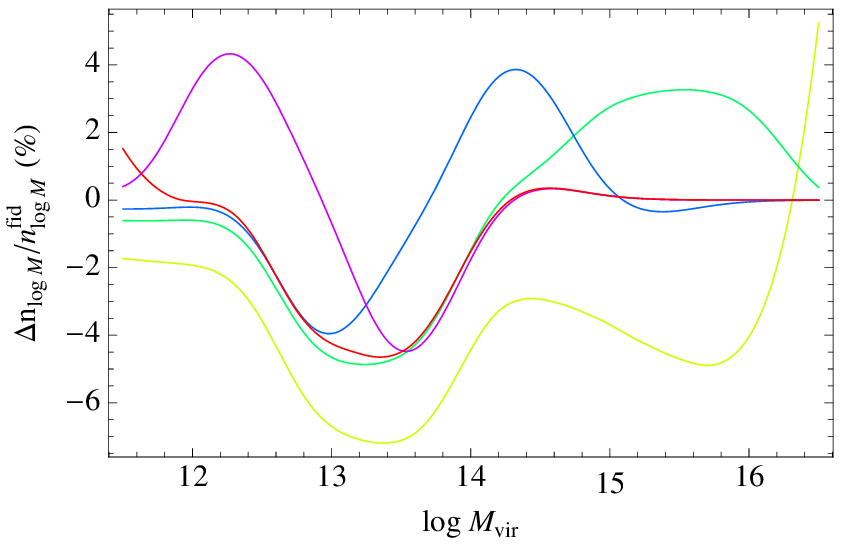}} \goodgap
\subfigure{\includegraphics[width=7.5cm]{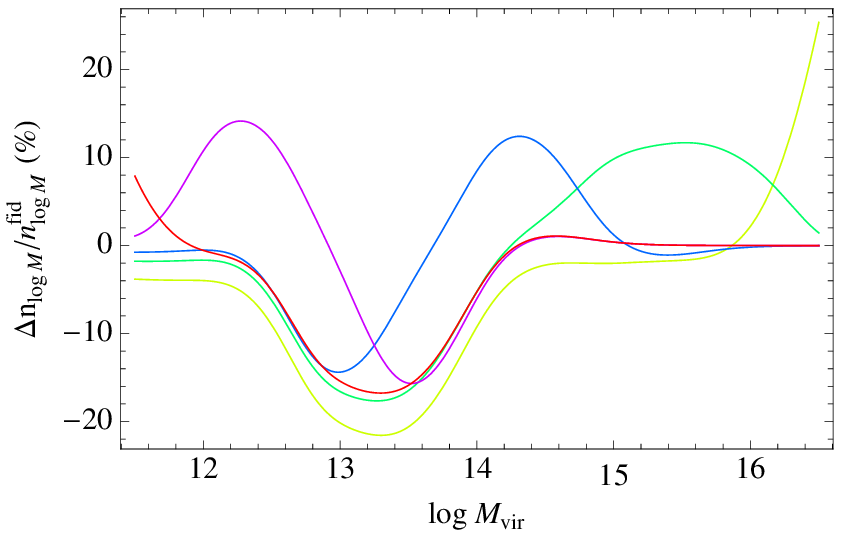}} \goodgap \\
\subfigure{\includegraphics[width=7.5cm]{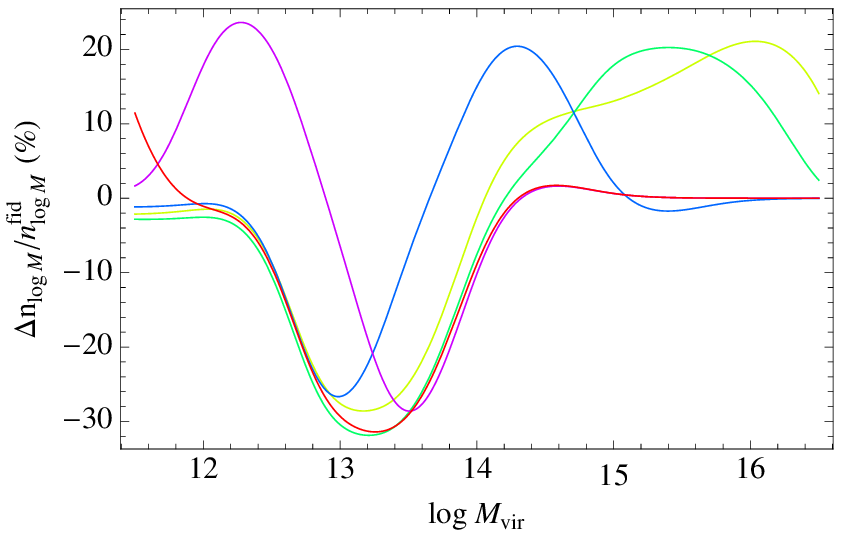}} \goodgap
\subfigure{\includegraphics[width=7.5cm]{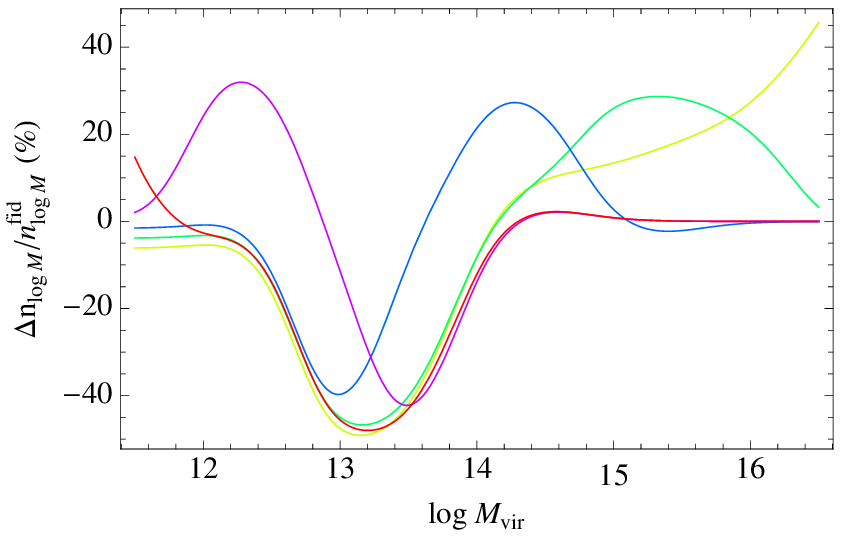}} \goodgap
\caption{Pecentage deviation, $\Delta {\cal{N}}(\log{M_{vir}})/{\cal{N}}(\log{M_{vir}})$ as a function of the halo mass for $z$ from 0.2 to 1.4 in steps of 0.4 (from top left to bottom right) and $\varepsilon$ from -3.5 to -7.5 in steps of -1.0 (yellow, green, blue, purple, red lines, respectively).}
\label{fig: mfhspar}
\end{figure*}

\subsection{Fiducial model and theoretical mass function}

The approximated formula (\ref{eq: hubapprox}) for $E(z)$ and the \cite{LH11} formalism allows us to estimate the mass function for the HS $f(R)$ model provided we set the parameters $(\Omega_M, h, q_0, j_0, n, \varepsilon)$. Moreover, one also has to set the slope of the power spectrum through $n_{PS}$ and fix its normalization through $\sigma_8$. Since we are mainly interested in how the mass function looks in $f(R)$ models and whether it is possible to discriminate among the HS model and its corresponding GR based dark energy model with the same expansion rate, we will not vary all the parameters at play. On the contrary, we will first fix the expansion rate to that of a CPL model with

\begin{displaymath}
(\Omega_M, h, w_0, w_a) = (0.273, 0.703, -0.95, 0.0) \ .
\end{displaymath}
Second, we fix the shape and the amplitude of the matter power spectrum setting $(n_{PS}, \sigma_8) = (0.966, 0.809)$ in agreement with \cite{WMAP7} constraints. As a fiducial HS model, we choose the one obtained by setting $(n, \varepsilon) = (1.5, -6.0)$ since it best mimics the CPL expansion rate. Moreover, this model is quite similar to the best fit one found by \cite{CCD11} fitting a large dataset including SNeIa, GRBs, BAOs and WMAP7 distance priors.

The mass function (MF) for this parameter set (which we will refer to in the following as the fiducial model) is plotted in the left panel of Fig.\,\ref{fig: mffid} for three different redshift values.  Although not clearly visible, a kink is present in the MF at the scale $\log{M_{vir}} \sim 13$ as a consequence of the interpolation scheme adopted to reproduce the MF from N\,-\,body simulations. Actually, such a feature is not an artifact, but rather the evidence of the chameleon effect. Indeed, the scale at which the kink is present is the same as the one marking the onset of the chameleon mechanism which increases the abundance of intermediate mass haloes. This can also be seen from the right panel where we compare the HS MF with the CPL one. For $M_{vir} >> M_{th}$, the chamaleon effect fully masks the enhancement of the gravitational force introduced by the modified gravity potential thus leading to a MF which is the same as the CPL one. In the opposite regime, the effective field is no more negligible and the gravitational force is significantly boosted leading to a marked increase of the abundance of intermediate mass haloes. Since the strength of the effective field is a function of $z$, the threshold chamaleon mass will change with the redshift so that the ratio between the HS and CPL MFs does depend not only on the mass, but on the redshit too.

Although discriminating between HS and CPL is interesting in its own, it is also instructive to look at how the HS MF depend on the $f(R)$ parameters. This can be inferred from Fig.\,\ref{fig: mfhspar} where we plot the percentage deviation from the fiducial model varying $\varepsilon$ for the fiducial $n$ at different redshifts. As an encouraging result, we find that $\Delta {\cal{N}}/{\cal{N}}$ can take quite large values in the range $12 \le \log{M_{vir}} \le 15$ for all $z$. Similarly large values can be achieved for larger masses, but we warn the reader not to overcome the results in this range since they refer to MFs which have almost vanished so that even a small difference in value gives rise to a large $\Delta {\cal{N}}/{\cal{N}}$. Explaining the shape of $\Delta {\cal{N}}/{\cal{N}}$ is not a straightforward task. On one hand, the scaling of the chameleon threshold mass $M_{th}$ as $|f_{R0} - 1|^{3/2} = {\rm dex}(3\varepsilon/2)$ implies that deviations from the fiducial MF are positive or negative depending on the mass being larger or smaller than the corresponding chamaleon mass. On the other hand, since ${\cal{G}}_{eff}$ also depends on $\varepsilon$, the mass variance $\sigma(M_{vir})$, which enters the \cite{ST99} profile through $\nu = \delta_c/\sigma$, will be a function of $\varepsilon$ thus introducing a second way through which $\varepsilon$ affects the MF. It is actually difficult to disentangle the two effects so that the final non monotonic behaviour in Fig.\,\ref{fig: mfhspar} is not immediate to quantitatively explain. It is, on the contrary, much easier to understand why we find that $n$ has only a very minor role in setting the shape and the amplitude of the MF with $\Delta {\cal{N}}/{\cal{N}}$ being smaller than $1\%$. Indeed, since the background expansion is almost the same for each $n$ (see, e.g., the discussion in \citealt{CCD11}), the only way $n$ can impact the MF is through the effective gravitational constant ${\cal{G}}_{eff}$. For the HS models we are considering, this dependence is actually quite mild and is further smoothed out by the procedure leading to the final MF. Note that this is partially a consequence of how we set the fiducial parameters. Indeed, for $(n, \varepsilon) = (1.5, -6.0)$, we get $c_1/c_2^2 \simeq 0.004$ so that, in the high curvature regime, we get $f(R) \sim R - m^2 c_1/c_2$ with $n$ playing a marginal role in setting the ratio $c_1/c_2$. Should we have chosen, e.g., $(n, \varepsilon) = (1.5, -4.0)$ as fiducial, we would have obtained $c_1/c_2^2 \simeq 0.4$ and hence the term $m^2 (c_1/c_2^2) (m^2/R)^n$ in the high curvature Lagrangian would have played a much significant role likely leading to a stronger dependence of the MF on $n$. However, such a large $\varepsilon$ does not lead to a background expansion in agreement with our fiducial CPL one so that we have not considered this kind of models.

As a final remark, we stress that both Figs.\,\ref{fig: mffid} and \ref{fig: mfhspar} refer to the theoretical MF. This is not what is actually observed so that the differences between the HS and CPL MFs or among the different HS model parameter sets can be strongly suppressed when the selection effects introduced by the particular method adopted to observationally determine the MF are taken into account. Should the observational MF be defined only for $\log{M_{vir}} > 14$ or be measured with too large uncertainties at higher $z$, the possibility to discriminate among the HS and CPL models and/or to constrain the HS parameters $(n, \varepsilon)$ would be seriously compromised.

\section{Weak lensing peaks}

Being the largest and most massive mass concentrations, galaxy clusters are ideal sites to look for gravitational lensing effects. Should the source be aligned with the cluster, the formation of spectacular arcs takes place, but, for most cases, the main effect is a coherent distortion of the shape of background galaxies, which appears stretched tangentially around the cluster. The distortion field can be used to reconstruct the shear map and then infer constraints on the cluster mass distribution. On the other hand, a shear map of the sky presents distinct peaks corresponding to the cluster positions thus offering a technique based on weak lensing to identify galaxy clusters.

As a peak finder, we consider here the aperture mass defined by \citep{S96}

\begin{equation}
M_{ap}(\theta) = \int{\kappa(\theta) U(\vartheta - \theta) d^2\theta}
= \int{\gamma_t(\theta) Q(\vartheta - \theta) d^2\theta}
\label{eq: mapdef}
\end{equation}
where $\kappa(\theta)$ and $\gamma_t(\theta) = -{\cal{R}}[\gamma(\theta) \exp{(-2 {\rm i} \phi)}]$ are the convergence and the tangential shear at position $\theta = (\vartheta \cos{\phi}, \vartheta \sin{\phi})$ and $U(\vartheta)$, $Q(\vartheta)$ are compensated filter functions related to each other by the integral equation

\begin{displaymath}
Q(\vartheta) = - U(\vartheta) + \frac{2}{\vartheta^2} \int_{0}^{\vartheta}{U(\vartheta^{\prime}) \vartheta^{\prime} d\vartheta^{\prime}} \ .
\end{displaymath}
In order to detect a cluster as a peak in the aperture mass map, we have to preliminarily estimate the $M_{ap}$ variance and then set a cut on the signal\,-\,to\,-\,noise (S/N) ratio. To this end, we have to specify how we compute both the signal and the noise which we address in the two following subsections.

\subsection{Halo model}

For given lens and source redshift $(z_l, z_s)$, the aperture mass depends on the mass density profile of the halo acting as a lens. We will model cluster haloes using the spherically symmetric NFW profile \citep{NFW97}\,:

\begin{equation}
\rho(r) = \frac{\rho_s}{x (1 + x)^2} = \frac{(M_{vir}/4 \pi R_{vir}^3) f_{NFW}(c_{vir})}{(c_{vir} y) (1 + c_{vir} y)^2}
\label{eq: rhonfw}
\end{equation}
with $x = r/R_s$, $y = r/R_{vir}$ and

\begin{equation}
f_{NFW}(c_{vir}) = \frac{c_{vir}^3}{\ln{(1 + c_{vir})} - c_{vir}/(1 + c_{vir})} \ .
\label{eq: fc}
\end{equation}
Note that we have reparameterized the model in terms of the virial mass $M_{vir}$, defined as the mass  within the virial radius $R_{vir}$ where the mean density equals $\Delta_{vir} \rho_{crit}(z_l)$ with $\rho_{crit}$ the critical density at the lens redshift, and the concentration $c_{vir} = R_{vir}/R_s$. According to N\,-\,body simulations, the NFW model can be reduced to a one parameter class since $c_{vir}$ correlates with the virial mass $M_{vir}$. Actually, the slope, the scatter and the redshift evolution of the $c_{vir}$\,-\,$M_{vir}$ relation are still matter of controversy with different results available in the literature. As a fiducial case, here we follow \cite{MC11} setting

\begin{eqnarray}
\log{c_{vir}} & = & (0.029 z_l - 0.097) \log{(h^{-1} M_{vir})} \nonumber \\
~ & - & \frac{110.001}{z_l + 16.8885} + \frac{2469.720}{(z_l + 16.8885)^2} \ ,
\label{eq: cvmvmc}
\end{eqnarray}
referring to this in the following as the MC11 relation. However, in order to investigate the  dependence of the results on the adopted mass\,-\,concentration law, we will also consider the $c_{vir}$\,-\,$M_{vir}$ relation empirically found by Buote et al. (2007, hereafter B07) from observed X\,-\,ray galaxy systems spanning the mass range $(0.06, 20) \times 10^{14} \ {\rm M}_{\odot}$, namely\,:

\begin{equation}
c_{vir} = \frac{c_0}{1 + z_l} \left ( \frac{M_{vir}}{10^{14} M_{\odot}} \right )^{\alpha}
\label{eq: cvmvbul}
\end{equation}
with $(c_0, \alpha) = (9.0, -0.172)$ as inferred from the fit to the full sample. It is worth noting that both $c_{vir}$\,-\,$M_{vir}$ relations are affected by a significant scatter. In order to take this into account, one should convolve the aperture mass for a given $M_{vir}$ value with a lognormal distribution centred on the predicted $c_{vir}$ value and with variance equal to the scatter itself. However, we follow here the common practice of neglecting this issue and simply insert the analytical expression for the convergence $\kappa$ of the NFW profile \citep{B96,WB00} into Eq.(\ref{eq: mapdef}) with $c_{vir}$ set according to Eq.(\ref{eq: cvmvmc}) or (\ref{eq: cvmvbul}).

A caveat is in order here. Both the NFW profile and the $c_{vir}$\,-\,$M_{vir}$ relations have been inferred from N\,-\,body simulations carried on in a $\Lambda$CDM cosmological framework hence implicitly assuming that, on galaxy scales, the gravitational potential is Newtonian. On the contrary, we are here investigating $f(R)$ theories so that, strictly speaking, the classical Newtonian theory does not hold anymore and the potential is modified. Actually, a Yukawa\,-\,like term is added to the Newtonian potential with a scale length which, due to the chameleon effect, depends on the environment and the strength of the effective field related to the modified gravity Lagrangian \citep{F07,CC11}. However, for the typical values of the $f(R)$ model parameters we are going to consider, the effective potential is almost identical to the Newtonian one so that we do not expect deviations of the halo profile from the NFW one. This is indeed what \cite{S09} have found in their analysis of the halo profiles from N\,-\,body simulations carried on for the same HS model we are considering here. Although a $c_{vir}$\,-\,$M_{vir}$ relation was not reported, we expect that the (unknown) actual relation is within the range set by the MC11 and B07 relations.

\subsection{Filter function and $S/N$ ratio}

It is worth noting that, being sensitive to all the matter along the line of sight, the observed $M_{ap}$ is actually the sum of the contribution due to both the cluster and the uncorrelated large scale structure projected along the same line of sight, i.e. $M_{ap} = M_{ap}^{clus} + M_{ap}^{LSS}$. In the usual approach, one considers that, being a density contrast, $M_{ap}^{LSS}$ averages to zero so that no bias is introduced in the $M_{ap}$ statistics. On the contrary, the LSS will contribute to the variance representing an additional source of noise \citep{H01}.  The filter functional form and its parameters are then chosen based on a comparison with simulated data and the characteristics of the survey (see, e.g, \citealt{H05}). Actually, such a method is not fully efficient in filtering out the LSS contribution and is, moreover, related to the underlying cosmology adopted in the reference simulation.

A possible way out of this problem relies on the use of the so called {\it optimal filter} introduced by Maturi et al. (2005, hereafter M05) to take explicitly into account both the shape of the halo signal and the underlying cosmology. Following M05, we set the Fourier transform of the filter as

\begin{equation}
\hat{\Psi}(\ell) = \frac{1}{(2 \pi)^2} \left [ \int{\frac{|\hat{\tau(\ell)}|^2}{P_N(\ell)} d^2 \ell} \right ]^{-1}
\frac{\hat{\tau}(\ell)}{P_N(\ell)} \ ,
\label{eq: defpsihat}
\end{equation}
where $\hat{\tau}$ is the Fourier transform of the signal (in our case, the tangential shear component) and $P_N(\ell)$ the noise power spectrum as a function of the angular wavenumber $\ell$. This latter is made up by the sum of two terms\,:

\begin{equation}
P_N(\ell) = P_{\varepsilon} + P_{\gamma}(\ell) \ ,
\label{eq: defpntot}
\end{equation}
with

\begin{equation}
P_{\varepsilon} = \frac{1}{2} \frac{\sigma_{\varepsilon}^2}{n_g}
\label{eq: defpeps}
\end{equation}
the noise contribution from the finite number of galaxies (with number density $n_g$) and their intrinsic ellipticities ($\sigma_{\varepsilon}$ being the variance), and $P_{\gamma}(\ell) = \frac{1}{2} P_{\kappa}(\ell)$ the noise due to the LSS. Note that this latter is set equal to half the convergence power spectrum since we use only one component of the shear. Under the Limber flat sky approximation, it is\footnote{Eq.(\ref{eq: limber}) is the same as in GR based dark energy models which could be surprising at first sight since we are using a modified gravity theory. Actually, it has been shown \citep{T07} that, for $f(R)$ theories, the convergence power spectrum is still given by Eq.(\ref{eq: limber}) provided the matter power spectrum $P_{\delta}(\ell/\chi, z)$ is computed taking care of the running effective gravitational constant and the result is multiplied by the factor $1/f^{\prime}(R)$. Since, for the models we are considering, $f^{\prime}(R) \simeq 1$ with great accuracy, we have neglected this correction when computing $P_{\kappa}$.}

\begin{equation}
P_{\kappa}(\ell) =
\left ( \frac{3 \Omega_M H_0^2}{2 c^2} \right )^2 \int_{0}^{\chi_h}{P_{\delta}\left ( \frac{\ell}{\chi}, \chi \right ) \frac{{\cal{W}}^2(\chi)}{a^2(\chi)} d\chi}
\label{eq: limber}
\end{equation}
with

\begin{equation}
\chi(z) = \frac{c}{H_0} \int_{0}^{z}{\frac{dz^{\prime}}{E(z^{\prime})}}
\label{eq: defchi}
\end{equation}
the comoving distance to redshift $z$ (with $\chi_h$ the distance to the last scattering surface), ${\cal{W}}(\chi)$ is the lensing weight fucntion

\begin{equation}
{\cal{W}}(\chi) = \int_{\chi}^{\chi_h}{\left ( 1 - \frac{\chi}{\chi^{\prime}} \right ) p_{\chi}(\chi^{\prime}) \chi^{\prime} d\chi^{\prime}}
\label{eq: lensweight}
\end{equation}
and $p_{\chi}(\chi) d\chi = p_{z}(z) dz$ is the source redshift distribution here parameterized by \citep{M11}

\begin{equation}
p_z(z) = \frac{\beta}{z_0} \left ( \frac{z}{z_0} \right )^2 \exp{\left [ - \left ( \frac{z}{z_0} \right )^{\beta} \right ]}
\label{eq: pz}
\end{equation}
and normalized to unity. Following \cite{M11}, we set $(\beta, z_0) = (1.5, 0.6)$ so that $z_m = 0.9$ is the median redshift of the sources as expected for the Euclid mission.

Following \cite{B96} and \cite{WB00} for the shear profile of the NFW model, one finally gets for the Fourier transform of the filter\footnote{{\bf Note that the minus sign comes out from our convention on the sign of the tangential shear component.}}\,:

\begin{eqnarray}
\hat{\Psi}(\ell) & = & - \frac{1}{(2 \pi)^3} \left [ \left (  \frac{M_{vir}/4 \pi R_{vir}^2}{\Sigma_{crit}} \right ) g(c_{vir})
\left ( 2 \pi \theta_s^2 \right ) \right ]^{-1} \nonumber \\
 & \times & \frac{\tilde{\tau}(\ell \theta_s)}{P_N(\ell) \tilde{{\cal{D}}}(\theta_s)}
\label{eq: fourierpsi}
\end{eqnarray}
where $\Sigma_{crit} = c^2 D_s/(4 \pi G D_d D_{ds})$ is the critical density for lensing (depending on the lens and source redshift), $g(c_{vir}) = f_{NFW}(c_{vir})/c_{vir}$,  $\theta_s$ is the angular scale corresponding to the characteristics radius $R_s$, and we have defined the dimensionless quantities

\begin{equation}
\tilde{\tau}(\ell \theta_s) = \int_{0}^{\infty}{\tilde{\gamma}(\xi) J_2(\ell \theta_s \xi) \xi d\xi} \ ,
\label{eq: deftautilde}
\end{equation}

\begin{equation}
\tilde{{\cal{D}}}(\theta_s) = \int_{0}^{\infty}{\frac{\left | \tilde{\tau}(\ell \theta_s) \right |^2}{P_N(\ell)} \ell d\ell} \ ,
\label{eq: defdtilde}
\end{equation}
with $\tilde{\gamma}(\xi)$ the dimensionless shear profile given by \citep{WB00}. Taking the {\bf back} Fourier transform of Eq.(\ref{eq: fourierpsi}) and setting $Q = \Psi$ in the aperture mass definition, it is only a matter of algebra to finally get\,:

\begin{equation}
M_{ap}(\vartheta; \theta_s) = \frac{1}{(2 \pi)^4} \frac{\tilde{M}_{ap}(\vartheta, \theta_s)}{\tilde{{\cal{D}}}(\theta_s)}
\label{eq: endmap}
\end{equation}
with ($\xi = \vartheta/\theta_s$ and $\xi^{\prime} = \theta/\theta_s$)

\begin{eqnarray}
\tilde{M}_{ap} & = & \int_{0}^{\xi}{\tilde{\gamma}(\xi^{\prime}) \xi^{\prime} d\xi^{\prime}} \\
 &  &
\int_{0}^{2 \pi}{\tilde{\Psi}\left [ \theta_s \left ( \xi^2 + \xi^{\prime 2} - 2 \xi \xi^{\prime} \cos{\theta} \right )^{1/2} \right ]
\cos{(2 \theta)} d\theta} \ . \nonumber
\label{eq: defmaptilde}
\end{eqnarray}
{\bf A conceptual remark is in order here. Eqs.(\ref{eq: endmap}) has been obtained starting from Eq.(\ref{eq: mapdef}) and assuming that the shear profile is the same as the cluster one. Actually, the tangential shear in Eq.(\ref{eq: mapdef}) is obtained from the observed galaxies ellipticities so that it is not strictly equal to the model shear which one estimates from the cluster model. However, since we are interested in making analytic predictions, such an identification is unavoidable. It is also worth noting that what is actually observable is the reduced shear $\gamma/(1 - \kappa)$ which is, however, equal to $\gamma$ in the weak lensing limit $(\kappa << 1)$ we are interested in here.}

A naive look at Eqs.(\ref{eq: endmap})\,-\,(35) could lead to the surprising conclusion that the aperture mass does not depend on the halo properties. This is, of course, not the case. First, the halo virial mass $M_{vir}$ and redshift $z_l$ directly enter the definition of $\theta_s$ which also depends on the adopted $c_{vir}$\,-\,$M_{vir}$ relation. Second, the halo model sets the dimensionless shear profile $\tilde{\gamma}(\xi)$ which enters both in the filter derivation and the $\tilde{M}_{ap}$ function. On the contrary, the dependence on the source redshift $z_s$ has been integrated out when computing the filter function. This is a consequence of how the optimal filter has been constructed, i.e., imposing that the filter is optimized to find halos with a given mass profile, virial mass and redsift notwithstanding the source redshift.

This latter quantity, however, plays a role when computing the noise given by\footnote{{\bf Note that it is customary to use $P_{N}(\ell)$ instead of $P_{\varepsilon}$ in Eq.(\ref{eq: defsigmaap}) for the total variance to take into account the contribution of both the shot noise and the large scale structure. However, we are here considering the signal as the sum of the cluster and LSS peaks so that only $P_{\varepsilon}$ has to be considered as noise term.}} \citep{M05}

\begin{eqnarray}
\sigma_{ap}^2 & = & \frac{1}{2 \pi} \int_{0}^{\infty}{P_{\varepsilon} \left | \tilde{\Psi}(\ell) \right |^2 \ell d\ell} \nonumber \\
 & = & \frac{1}{(2 \pi)^7} \left ( \frac{M_{vir}/4 \pi R_{vir}^2}{\Sigma_{crit}} \right )^{-2} \frac{g^{-2}(c_{vir})}
{\left ( 2 \pi \theta_s^2 \right )^2} \frac{P_{\varepsilon}}{\tilde{{\cal{D}}}(\theta_s)} \nonumber \\
 & \times & \int_{0}^{\infty}{\frac{\tilde{\tau}^2(\ell \theta_s)}{[P_{\varepsilon} + (1/2) P_{\kappa}(\ell)]^2} \ell d\ell}
\label{eq: defsigmaap}
\end{eqnarray}
so that the $S/N$ ratio reads\footnote{{\bf A cautionary remark is in order here. The $(M_{vir}, c_{vir})$ parameters entering the $S/N$ ratio refer to the cluster halo responsible of the signal one is detecting. This halo can also be different from the template model used to generate the filter. However, the $(M_{vir}, c_{vir})$ template parameters setting the filter normalization cancels out when computing the $S/N$ ratio so that, in order to simplify the notation, we have not used different symbols to differentiate the template and actual halo parameters.}}

\begin{equation}
{\cal{S}}(\vartheta; {\bf p}) = \frac{1}{\sqrt{2 \pi}}  \frac{M_{vir}/4 \pi R_{vir}^2}{\Sigma_{crit}}
\frac{2 \pi \theta_s^2 g(c_{vir})}{\sqrt{P_{\varepsilon}}} \frac{\tilde{M}_{ap}(\vartheta)}{\tilde{\sigma}_{ap}}
\label{eq: stonvsz}
\end{equation}
where ${\bf p}$ summarizes the parameters the $S/N$ ratio depend on, namely the halo quantities $(M_{vir}, z_l)$ and the source redshift $z_s$ (entering because of $\Sigma_{crit}$). Since we are interested in the total $S/N$ ratio, we integrate over the source redshift distribution thus finally obtaining

\begin{equation}
{\cal{S}}(\vartheta; z_l, M_{vir}) = \int_{z_l}^{\infty}{{\cal{S}}(\vartheta; z_l, z_s, M_{vir}) p_z(z_s) dz_s}
\label{eq: stonend}
\end{equation}
which, for a given $c_{vir}$\,-\,$M_{vir}$ relation and mass profile, depends only on the halo virial mass and redshift.

\begin{figure*}
\centering
\subfigure{\includegraphics[width=5.25cm]{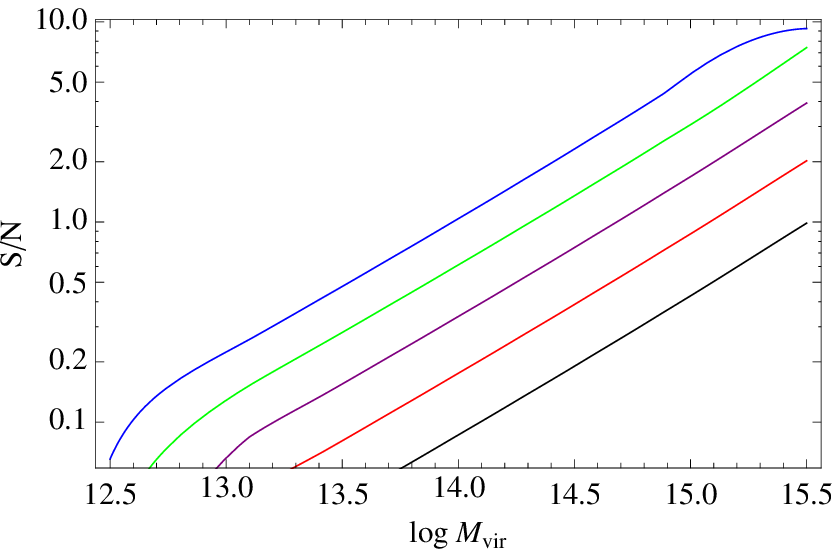}} \goodgap
\subfigure{\includegraphics[width=5.25cm]{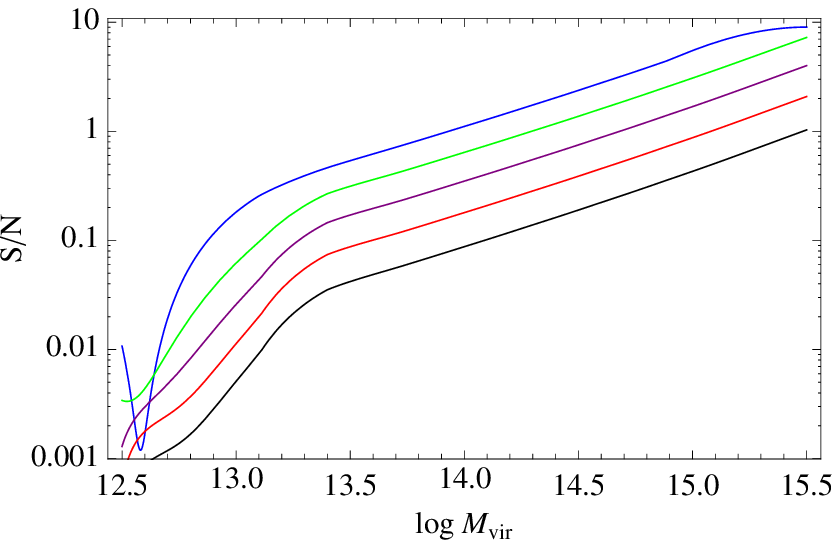}} \goodgap
\subfigure{\includegraphics[width=5.25cm]{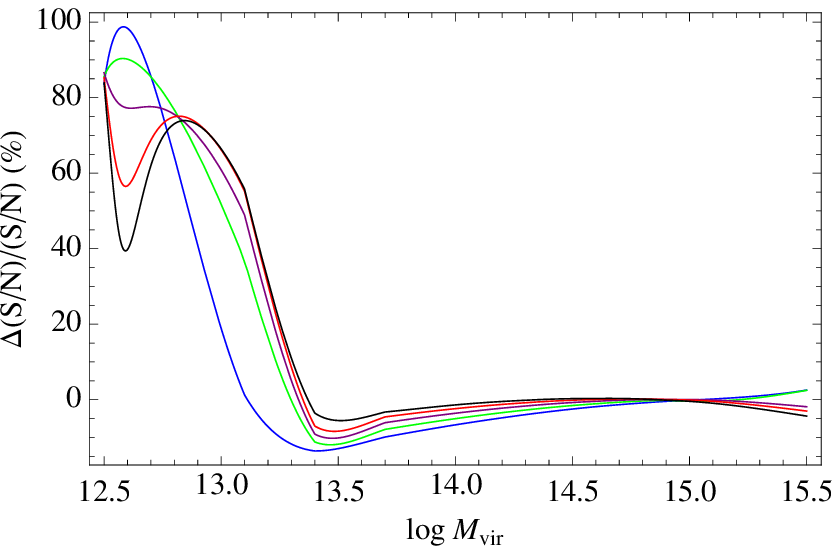}} \goodgap
\caption{{\it Left.} Signal\,-\,to\,-\,noise ratio as function of the virial mass for different redshift values, namely $z = 0.5, \ldots, 1.3$ in steps of 0.2 (from the upper blue to the lower black curves) adopting the fiducial cosmological model and the MC11 $c_{vir}$\,-\,$M_{vir}$ relation. {\it Centre.} Same as left panel but for the B07 $c_{vir}$\,-\,$M_{vir}$ relation. {\it Right.} $S/N$ percentage difference, $\Delta (S/N)/(S/N) = ({\cal{S}}_{MC11} - {\cal{S}}_{B07})/{\cal{S}}_{MC11}$, as a function of mass and redshift (same color scheme as before) for the fiducial cosmological model.}
\label{fig: snratio}
\end{figure*}

\subsection{Cluster detectability}

Eq.(\ref{eq: stonend}) enables us to estimate the $S/N$ ratio for a NFW halo with virial mass $M_{vir}$ and concentration $c_{vir}$ acting as a lens at redshift $z_l$. In order to compute ${\cal{S}}$, we need first to assign the survey characteristics (determining the noise power spectrum $P_{\varepsilon}$) and set the {\bf filter scale} $\vartheta$. We consider an Euclid\,-\,like survey\footnote{{\tt http://www.euclid-ec.org}} \citep{RB} and take a survey area of $15000 \ {\rm deg^2}$ with an ellipticity dispersion $\sigma_{\epsilon}= 0.3$ and a number of galaxies $n_g = 30 \ {\rm gal/arcmin}^2$. Note that these are the survey goal, but one can easily scale the $S/N$ ratio noting that ${\cal{S}} \propto \sqrt{n_g}$, while the total number of peaks scales linearly with the survey area.

The choice of the {\bf filter scale} $\vartheta$ asks for some caution. The optimal filter has been designed taking into account the NFW profile in order to maximize the signal. A natural scale would therefore be $\vartheta = \theta_s$ since most of the mass contributing to the lensing signal is contained within this aperture. For a cluster with $M_{vir} = 10^{15} \ {\rm M_{\odot}}$ at the median survey redshift $z_l = 0.9$, we get $\theta_s \simeq 1 \ {\rm arcmin}$, while the virial radius subtends an angle $\theta_{vir} \simeq 3.5 {\rm arcmin}$ (in our adopted fiducial cosmology). However, not all the clusters have the same mass and the same redshift. Setting $\vartheta = 1 \ {\rm arcmin}$ would be an optimal choice for these median values, but will strongly underestimate the signal for clusters at lower redshifts. On the contrary, a varying $\vartheta$ would allow to maximize the $S/N$ at every redshift, but would introduce an inhomogeneity in the computation of the number of peaks. As a compromise, we therefore set $\vartheta = 2 \ {\rm arcmin}$ noting that the filter actually cuts the contribution from regions outside few times $\theta_s$ and vanishes on scales larger than $\theta_{vir}$ (which can be smaller than $2 \ {\rm arcmin}$ for low mass and/or high redshift clusters).

It is instructive to look at the $S/N$ ratio as a function of the cluster redshift and virial mass for a given cosmological model. To this end, we consider the fiducial HS model with $(\Omega_M, n, \varepsilon) = (0.273, 1.5, -6.0)$ and look at how ${\cal{S}}$ depends on the adopted $c_{vir}$\,-\,$M_{vir}$ relation. Fig.\,\ref{fig: snratio} shows ${\cal{S}}$ as function of the virial mass at different redshift values for the MC11 and B07 cases (left and central panels, respectively). For both models, the trend with $(z, \log{M_{vir}})$ is the same\,: the higher the redshift, the larger must the virial mass be in order to attain a given ${\cal{S}}$ value. This can be qualitatively explained by noting that the integration in Eq.(\ref{eq: stonend}) is performed on a smaller interval with increasing $z$ thus reducing ${\cal{S}}$. In other words, the larger is the cluster redshift, the lower is number of sources available for lensing so that the $S/N$ ratio takes smaller values. In order to compensate for this reduction, one has to increase the virial mass since ${\cal{S}}$ approximately scales as $M_{vir}^{\alpha}$ with the slope $\alpha$ depending on the adopted $c_{vir}$\,-\,$M_{vir}$ relation.

The right panel in Fig.\,\ref{fig: snratio} compares ${\cal{S}}$ for the MC11 and B01 cases as a function of $M_{vir}$ and $z_l$ showing that the $S/N$ ratio turns out to depend strongly on the $c_{vir}$\,-\,$M_{vir}$ relation. In particular, for less massive systems, the MC11 relation predicts a values of ${\cal{S}}$ which can be up to $100\%$ larger than the B07 ones. The difference is, however, not monotonic reverting its sign for intermediate mass haloes and becoming quite small for very large mass. Such a strong sensibility to the $c_{vir}$\,-\,$M_{vir}$ model is not unexpected and is a consequence of the different concentrations for the same virial mass. In particular, the ratio $c_{vir}^{B07}/c_{vir}^{MC11}$ depends on both $z_l$ and $M_{vir}$ thus explaining why the $\Delta {\cal{S}}/{\cal{S}}$ may change sign according to which model predicts the larger $c_{vir}$ value.

As a final remark, we stress that the above results are almost independent of the underlying cosmological model. This can be easily explained by looking at how cosmology enters in the evaluation of the S/N ratio. First, we note that ${\cal{S}}$ depends on the integrated Hubble rate through the lensing critical density $\Sigma_{crit}$. This quantity only weakly depends on cosmology since it involves an integral of a ratio of distances over the source redshift. On the other hand, the convergence power spectrum $P_{\kappa}(\ell)$ enters the filter definition and hence both $M_{ap}$ and $\sigma_{ap}$ so that its impact is reduced when considering the ratio of these two quantities. Moreover, for the HS model parameter space we are interested in, the Hubble rate $H(z)$ is almost the same as the CPL one so that changing the $(\Omega_M, n, \varepsilon)$ values has a very minor impact on the S/N ratio. This is indeed what we have checked. Varying $(\Omega_M, \varepsilon)$ shifts ${\cal{S}}$ by less than $10\%$ (but typically a shift larger than $1\%$ is difficult to get), while the $S/N$ ratio may be considered independent of $n$ within a very good approximation.

\begin{figure*}
\centering
\subfigure{\includegraphics[width=5.25cm]{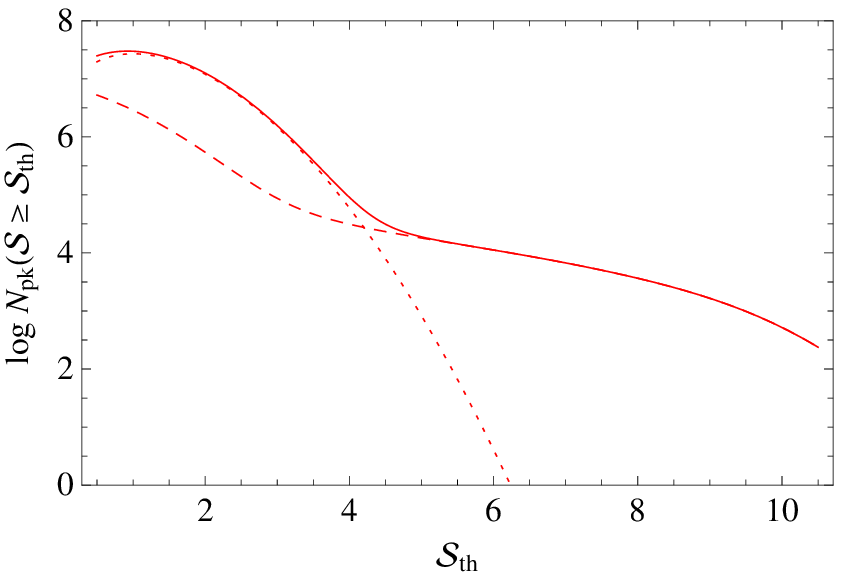}} \goodgap
\subfigure{\includegraphics[width=5.25cm]{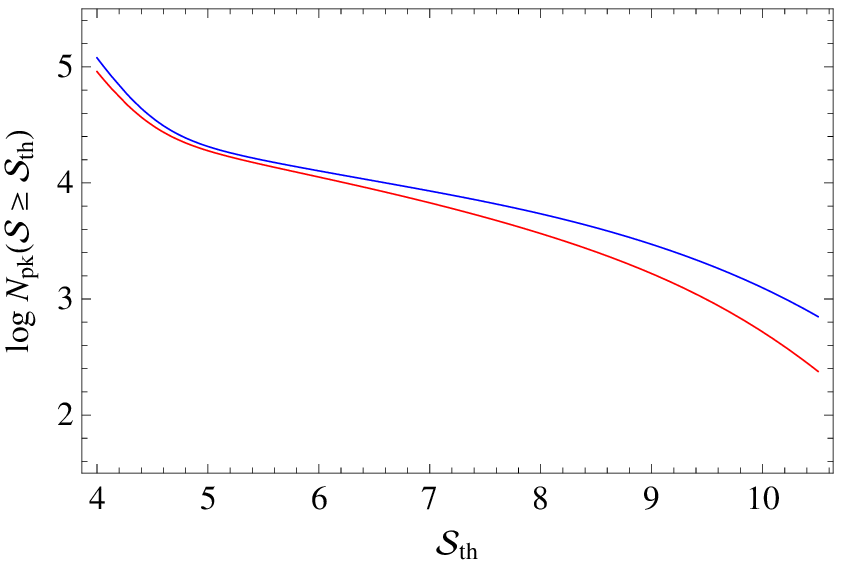}} \goodgap
\subfigure{\includegraphics[width=5.25cm]{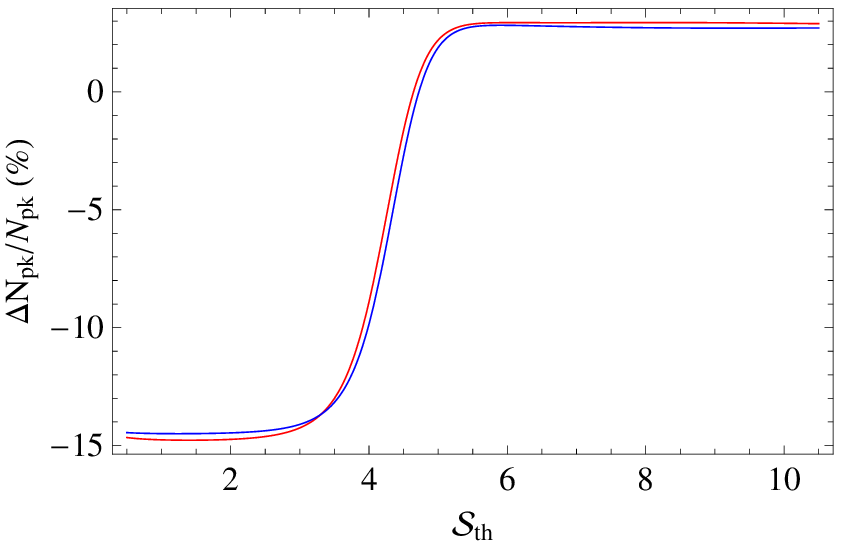}} \goodgap
\caption{{\it Left.} Total number of detectable peaks as a function of the threshold $S/N$ ratio for the fiducial HS model adopting the MC11 $c_{vir}$\,-\,$M_{vir}$ relation; solid, dashed and dotted lines refer to the total number and the contribution from clusters and LSS, respectively. {\it Centre.} Same as before but showing the total number only (in the regime dominated by the clusters contribution) for the MC11 (red) and B07 (blue) mass\,-\,concentration relations. {\it Right.} Percentage deviation, $\Delta N_{pk}/N_{pk} = (N_{pk}^{HS} - N_{pk}^{CPL})/N_{pk}^{HS}$ as a function of ${\cal{S}}_{th}$ with the labels HS and CPL referring to the fiducial HS and CPL models.}
\label{fig: npkfid}
\end{figure*}

\section{Peak number counts}

Having detailed how the S/N ratio can be computed and having determined the theoretical MF, we can now estimate the number density of haloes with mass $M_{vir}$ that produces significant peaks in the aperture mass map. To this end, we first have to take into account that a S/N threshold for the weak lensing signal does not correspond to an equally sharp threshold in halo mass because of the scatter in $M_{ap}$ caused by the shot noise from discrete background galaxy positions and the intrinsic ellipticity distribution. A halo of mass $M_{vir}$ has therefore a certain probability $p(M_{ap} | M_{vir})$ to produce an aperture mass $M_{ap}$ which we can model as a Gaussian\,:

\begin{equation}
p(M_{ap} | M_{vir}) \propto \exp{\left \{ - \frac{1}{2} \left [ \frac{M_{ap} - \hat{M}_{ap}(M_{vir})}{\sigma_{ap}} \right ]^2 \right \}}
\label{eq: mapprob}
\end{equation}
where $\hat{M}_{ap}(M_{vir})$ is the theoretically expected value and $\sigma_{ap}$ the variance in Eq.(\ref{eq: defsigmaap}) integrated over the source distribution. The probability that the $S/N$ ratio will be larger than a given threshold value will then be given by \citep{BPB02}\,:

\begin{equation}
p({\cal{S}} > {\cal{S}}_{th} | M_{vir}, z) = \frac{1}{2} {\rm erfc}\left [ \frac{{\cal{S}}(M_{vir}, z) - {\cal{S}}_{th}}{\sqrt{2}} \right ]
\label{eq: snrprob}
\end{equation}
so that the number density of haloes giving a detectable weak lensing peak finally reads

\begin{equation}
{\cal{N}}_{lens}(M_{vir}, z) = p({\cal{S}} > {\cal{S}}_{th} | M_{vir}, z) {\cal{N}}(M_{vir}, z)
\label{eq: nlens}
\end{equation}
where, hereafter, we will drop the label $l$ from $z$ to denote the cluster redshift. The total number of peaks generated by cluster haloes and with $S/N$ larger than a threshold value ${\cal{S}}_{th}$ is then obtained by integrating over $z$ and multiplying for the survey area $\Omega$ thus finally reading\,:

\begin{eqnarray}
N_{halo}({\cal{S}} > {\cal{S}}_{th}) & = & \left ( \frac{c}{H_0} \right )^3 \left ( \frac{\pi}{180} \right )^2 \left ( \frac{\Omega}{1 \ {\rm deg}^2} \right ) \\
~ & \times &  \int_{z_{L}}^{z_{U}}{\frac{r^2(z)}{E(z)} dz \nonumber
\int_{0}^{\infty}{{\cal{N}}_{lens}(M_{vir}, z) dM_{vir}}}
\label{eq: npktot}
\end{eqnarray}
where we set $(z_{L}, z_{U}) = (0.1, 1.4)$ as redshift limits\footnote{Note that the survey will likely detect galaxies over a much larger range, but we have cut the redshift range since, as will be shown later, the number of peaks outside this range is negligible.}, while $r(z) = (c/H_0)^{-1} \chi(z)$ is the comoving distance.

The number of observed peaks is, however, larger than ${\cal{N}}_{halo}$ because of the contamination from the LSS. This latter term may be estimated as \citep{M10,MFM11}\,

\begin{equation}
{\cal{N}}_{LSS} = \frac{1}{(2 \pi)^{3/2}} \left ( \frac{\sigma_{LSS}}{\sigma_{ap}} \right )^2
\frac{\kappa_{th}}{\sigma_{ap}} \exp{\left [ - \frac{1}{2} \left ( \frac{\kappa_{th}}{\sigma_{ap}} \right )^2 \right ]}
\label{eq: npklss}
\end{equation}
with $\kappa_{th} = {\cal{S}}_{th} \sigma_{ap}$ and

\begin{equation}
\left ( \frac{\sigma_{LSS}}{\sigma_{ap}} \right )^2 = \frac{\int_{0}^{\infty}{P_N(\ell) \left | \hat{\Psi}(\ell) \right |^2 \ell^3 d\ell}}
{\int_{0}^{\infty}{P_{\varepsilon}(\ell) \left | \hat{\Psi}(\ell) \right |^2 \ell d\ell}} \ .
\label{eq: defsigmaratio}
\end{equation}
Note that ${\cal{N}}_{LSS}$ only depends on the noise properties and the threshold $S/N$ ratio, but not on the lens mass and redshift since it is not related to the particular cluster at hand, but to the LSS. For this same reason, ${\cal{N}}_{LSS}$ is determined by the matter power spectrum (and hence the underlying cosmological scenario) entering $P_{N}(\ell)$. This will actually offers a further opportunity to discriminate among GR based dark energy models and modified gravity theories. The total number of observed peaks will finally be given by\,:
\begin{displaymath}
{\cal{N}}_{pk}({\cal{S}} > {\cal{S}}_{th}) = {\cal{N}}_{halo}({\cal{S}} > {\cal{S}}_{th}) +{\cal{N}}_{LSS}({\cal{S}}_{th}) \ .
\end{displaymath}

\begin{figure*}
\centering
\subfigure{\includegraphics[width=7.5cm]{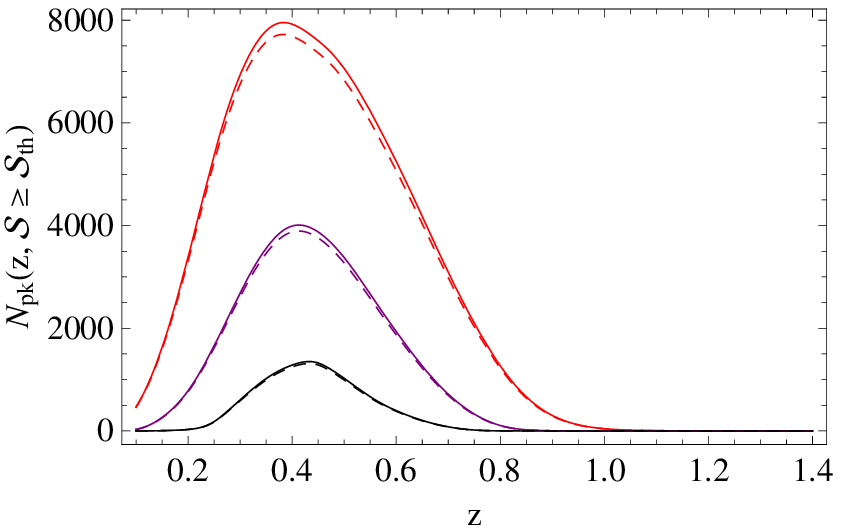}} \goodgap
\subfigure{\includegraphics[width=7.5cm]{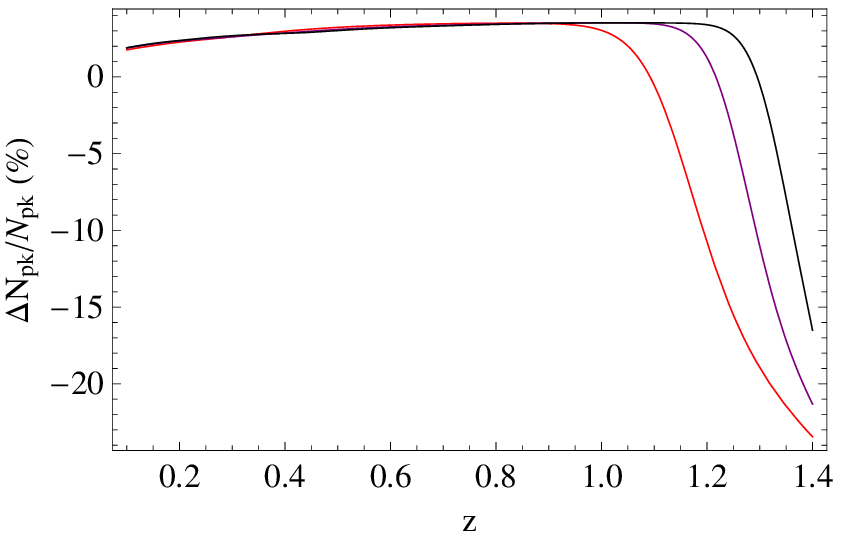}} \goodgap
\caption{{\it Left.} Total number of peaks in redshift bins for the HS (solid) and CPL (dashed) models adopting the MC11 mass\,-\,concentration relation and threshold $S/N$ ratio, ${\cal{S}}_{th} = 5, 7, 9$ (red, purple, black lines). {\it Right.} Percentage difference as a function of $z$ for the three ${\cal{S}}_{th}$ values adopted.}
\label{fig: npkvsz}
\end{figure*}

\begin{figure*}
\centering
\subfigure{\includegraphics[width=7.5cm]{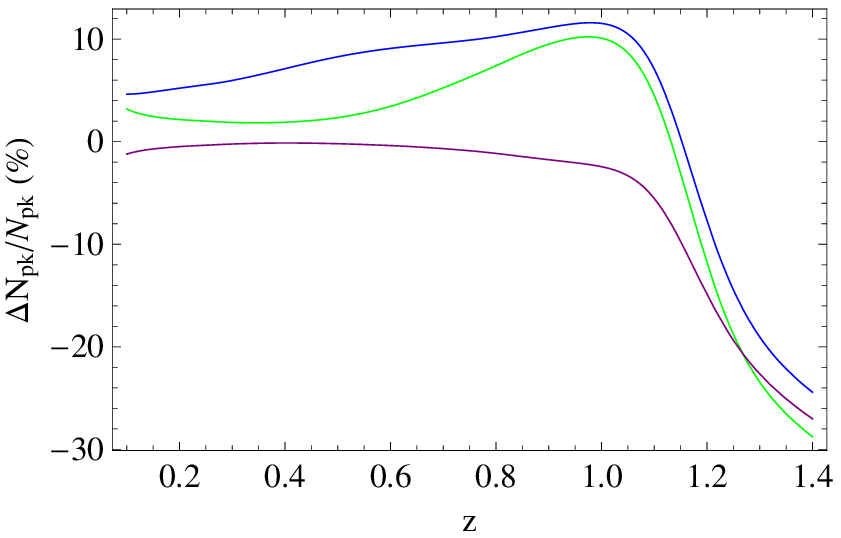}} \goodgap
\subfigure{\includegraphics[width=7.5cm]{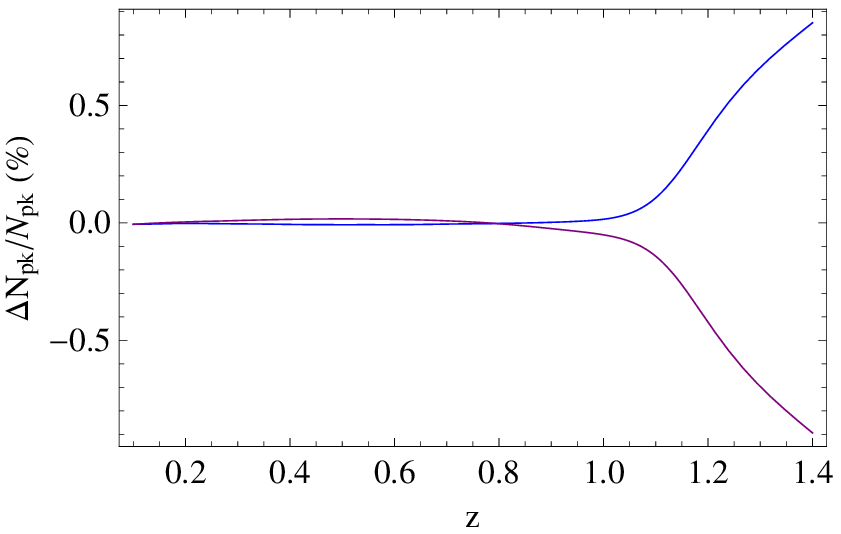}} \goodgap \\
\subfigure{\includegraphics[width=7.5cm]{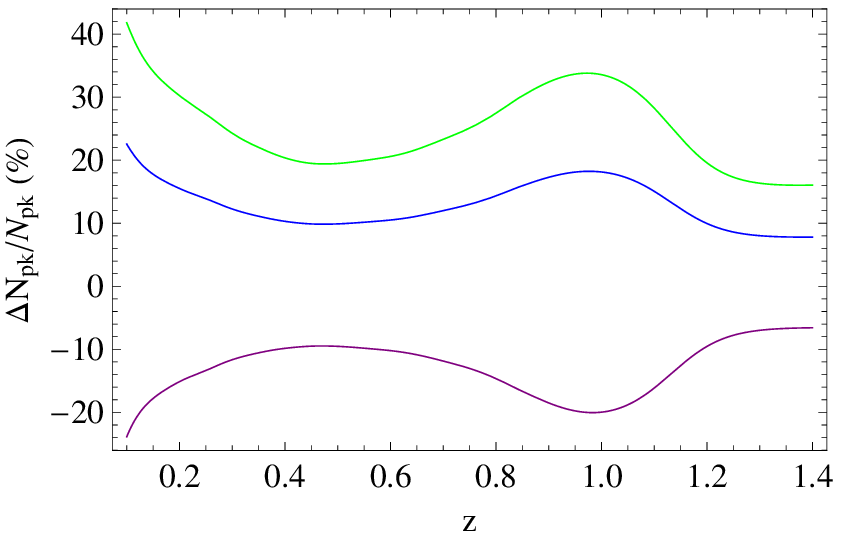}} \goodgap
\subfigure{\includegraphics[width=7.5cm]{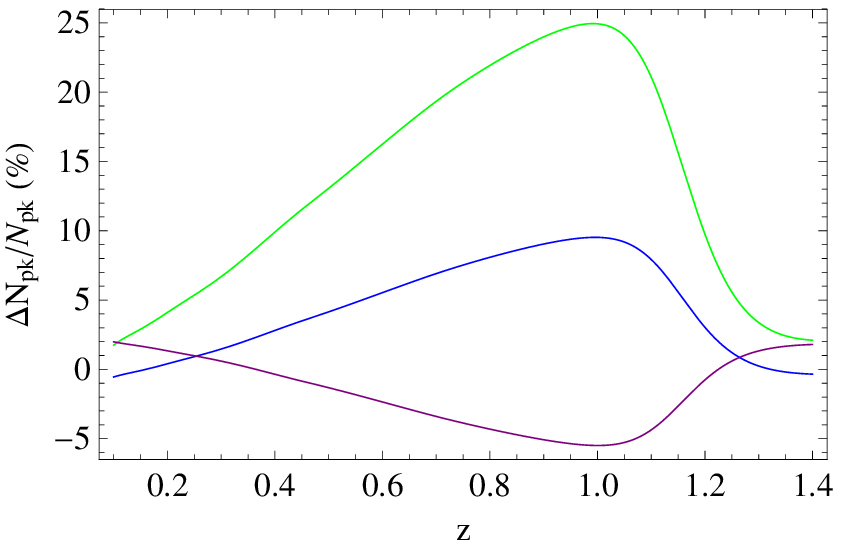}} \goodgap
\caption{Percentage deviation of $N_{pk}(z)$ as a function of $z$ with respect to the fiducial HS model with the MC11 $c_{vir}$\,-\,$M_{vir}$ relation. In each panel, only one of the parameters $(n, \varepsilon, \Omega_M, \sigma_8)$ is changed. {\it Top left.} Results for $\varepsilon = -4.0$ (green), $\varepsilon = -5.0$ (blue), $\varepsilon = -7.0$ (purple). {\it Top right.} Results for $n = 1.0$ (blue) and $n = 2.0$ (purple). {\it Bottom left.} Results for $\Omega_M = 0.234$ (green), $\Omega_M = 0.254$ (blue), $\Omega_M = 0.294$ (purple). {\it Bottom right.} Results for $\sigma_8 = 0.61$ (green), $\sigma_8 = 0.71$ (blue), $\sigma_8 = 0.91$ (purple).}
\label{fig: npkpar}
\end{figure*}

\noindent A look at Fig.\,\ref{fig: npkfid} helps us to highlight many important issues. First, in the left panel, we plot $N_{pk}({\cal{S}} > {\cal{S}}_{th})$ as a function of the threshold $S/N$ ratio for the fiducial HS model and the MC11 $c_{vir}$\,-\,$M_{vir}$ relation showing separately the contribution of the clusters and LSS terms. As expected, $N_{pk}$ is dominated by the LSS term for small $S/N$ ratio, that is to say, the smaller is the $S/N$ ratio, the larger is the probability that the detected peak is a fake due to the LSS rather than the evidence for a cluster. This is in agreement with common sense expectation and previous analysis in literature using different cosmological models and survey parameters \citep{H05,M10}. The term ${\cal{N}}_{LSS}$, however, quickly becomes subdominant so that one can confidently be sure that all peaks detected with a $S/N$ ratio larger than ${\cal{S}}_{th} \simeq 5$ are due to clusters. Note that we find a value for ${\cal{S}}_{th}$ comparable but larger than what is suggested in \cite{M10} because of differences in both the cosmological model and the survey characteristics.

The central panel in Fig.\,\ref{fig: npkfid} shows the impact of the adopted $c_{vir}$\,-\,$M_{vir}$ relation zooming on the high ${\cal{S}}$ region which is dominated by the clusters term. It turns out that the number of peaks is systematically larger if one adopts the B07 rather than the MC11 relation. This is actually not a fully unexpcted result. Indeed, as the right panel of Fig.\,\ref{fig: snratio} shows, the $S/N$ ratio is larger for the B07 model in the high mass regime. Put in other words, for a given ${\cal{S}}_{th}$, the number of high mass clusters with ${\cal{S}} > {\cal{S}}_{th}$ is larger for the B07 $c_{vir}$\,-\,$M_{vir}$ relation. Since these are the main responsible of the total number of high $S/N$ peaks, it is then straightforward to explain why we indeed find that $N_{pk}$ is larger for the B07 relation and why the offset increases with ${\cal{S}}$.

The most interesting issue to address is whether the number count of detectable peaks in the aperture mass maps allows to discriminate between the HS and CPL models notwithstanding their equal expansion rate. The right panel of Fig.\,\ref{fig: npkfid} offers a first answer to this question. Should we rely on clusters peaks setting ${\cal{S}} > 5$ for detection \citep{G10,K12}, the difference between the two models amounts to a modest $3\%$ with the HS case giving a larger number of peaks. This is somewhat surprisingly considering that the enhanced growth of structures in $f(R)$ theories should lead to a larger number of clusters for a given mass. Actually, such a result is not fully unexpected if one looks back at the right panel in Fig.\,\ref{fig: mffid}. Here, we have shown that the HS MF is larger than the CPL one for $M_{vir} < 10^{14} \ {\rm M}_{\odot}$, while the difference becomes negligible at higher masses because of the onset of the chameleon effect. Since the $S/N$ ratio for haloes with masses below $10^{14} \ {\rm M}_{\odot}$ is quite small (unless they are at low $z$), their contribute to the total number of peaks is strongly suppressed thus lowering the difference with the offset between the HS and CPL predictions. It is, however, worth noticing that this same offset significantly increases (up to $\sim 15\%$) and changes sign if we decrease the threshold $S/N$ ratio. This is related to the onset of the LSS dominated regime. In such a case, what is important is the difference in the convergence power spectrum entering the ratio $\sigma_{LSS}/\sigma_{ap}$, while $P_{\kappa}(\ell)$ plays a minor role in ${\cal{N}}_{lens}$. We can therefore conclude that the number of low $S/N$ peaks is a valuable tool to discriminate between HS and CPL models. It is also worth stressing that the offset between the two models predictions is (within a good approximation) independent of the adopted $c_{vir}$\,-\,$M_{vir}$ relation as can be seen from the closeness of the red and blu curves in the figure. This is an expected result considering that this ingredient only enters the determination of the $S/N$, but does not change how the number of peaks depend on the underlying cosmological model.

The total number of peaks is evaluated by integrating over the full redshift range thus smoothing out the dependence of the MF on $z$. Actually, how the MF changes with $z$ depends on the underlying cosmological model so that it is worth investigating what can be learned by considering the number of peaks in redshift bins. This is still given by Eq.(\ref{eq: npktot}) provided one replaces $(z_{L}, z_{U})$ with $(z - \Delta z/2, z + \Delta z/2)$. We set $\Delta z = 0.1$ and show the resulting $N_{pk}(z, {\cal{S}} > {\cal{S}}_{th})$ in Fig.\,(\ref{fig: npkvsz}). Note that here we have turned off the LSS term. Since these peaks are fake detections due to the LSS, it is obvious that no redshift can be attributed to them. We can therefore assume that they have been manually deleted from the sample and simply set\footnote{Such an assumption is actually also mathematically motivated. Indeed, we are now considering $d{\cal{N}}_{pk}/dz$. Since ${\cal{N}}_{pk}$ is the sum of the two terms due to the haloes and the LSS, respectively, and the LSS one is independent on $z$, we trivially get $d{\cal{N}}_{pk}/dz = d{\cal{N}}_{halo}/dz + d{\cal{N}}_{LSS}/dz = d{\cal{N}}_{halo}/dz = {\cal{N}}_{halo}(z)$.} ${\cal{N}}_{pk}(z) = {\cal{N}}_{halo}(z)$. The left panel suggests that binning the data has not improved the constraints, but this is only part of the story. Indeed, the percentage difference between HS and CPL predictions are quite modest $(\sim {\rm few} \ \%)$ up to $z \sim 1$, but they become as large as $20\%$ in the highest redshift bins. Such a behaviour can be qualitatively explained looking back at the right panel of Fig.\,\ref{fig: mffid} showing that the larger is $z$, the larger is the difference between the HS and CPL MF. Probing the high\,-\,$z$ end of the ${\cal{N}}_{pk}(z)$ curve is therefore an indirect method to investigate the MF in the regime where it is most sensible to the underlying gravity theory. As a further remark, we also note that the percentage difference between HS and CPL predictions at a given $z$ is smaller for higher $S/N$ thresholds. This can be qualitatively explained noting that, in order to have a large ${\cal{S}}$ peak, the cluster must be very massive so that the larger is ${\cal{S}}_{th}$, the more we are investigating the high mass tail of the MF. As already stated before, the HS and CPL MFs become degenerate for $M _{vir} > 10^{14} \ {\rm M}_{\odot}$ so that a sample probing this range will be unable to discriminate between the two competing models.

Although discriminating between CPL and HS scenarios is likely the most intriguing issue to consider, it is nevertheless of deep interest to investigate how $N_{pk}(z)$ depends on the HS model parameters. To this end, we hereafter adopt the MC11 relation since this gives a lower number of detections thus providing conservative estimates . Fig.\,\ref{fig: npkpar} shows that only $\varepsilon$ has an important effect on the peak number with deviations which can be of order $10 - 30\%$ depending on the redshift bin considered. Comparing Fig.\,\ref{fig: npkpar} with Fig.\,\ref{fig: mfhspar} clearly shows the impact of moving from a theoretical to an observable quantity. Indeed, the deviations from the fiducial model are almost halved the expected values, because only the haloes with masses large enough to lead to a $S/N$ ratio larger than the threshold can now be detected thus removing the part of Fig.\,\ref{fig: mfhspar} with large deviations. As a consequence, while the possibility of discriminating among different $\varepsilon$ values is still significant, the dependence on $n$ almost disappears with deviations being smaller than $1\%$ and only present in the highest redshift bins. Finally, although not $f(R)$ parameters, we also plot the deviations from the fiducial model due to changes of the matter density $\Omega_M$ and the variance $\sigma_8$. The bottom left panel of Fig.\,\ref{fig: npkpar} shows that deviations as high as $\sim 40\%$ can be obtained by changing $\Omega_M$, but the way $\Delta N_{pk}/N_{pk}$ depends on $z$ is different from what happens when varying $\varepsilon$ thus suggesting that no strong degeneracy between these two parameters is present. Similarly, the lower right panel tells us that $\sigma_8$ also plays a major role in determining the peak number counts in the highest redshift bins thus suggesting that strong constraints could be put on this parameter.

\begin{figure*}
\centering
\includegraphics[width=15cm]{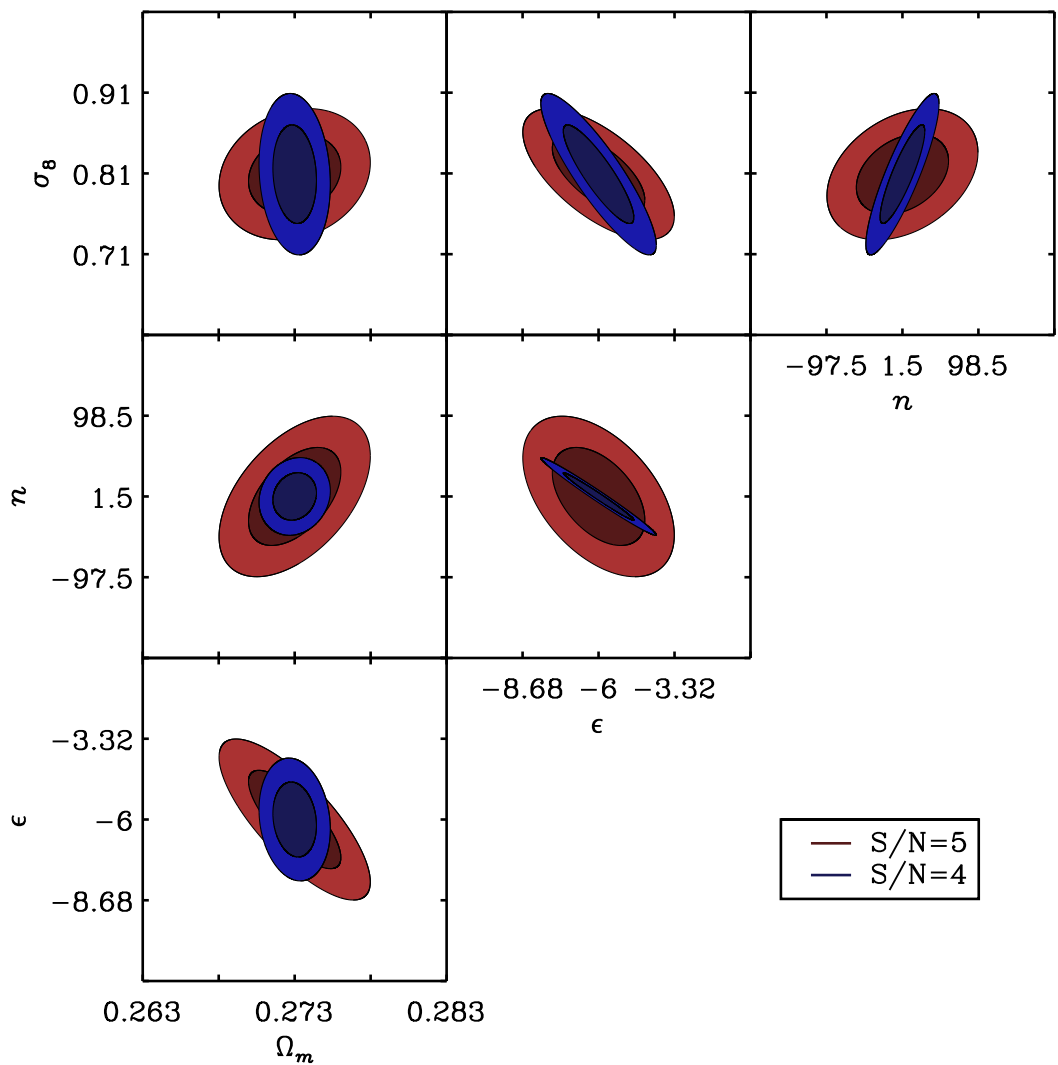}
\caption{Fisher matrix constrain in 2D spaces marginalizing over all the parameters but the ones on the axes.}
\label{fig: fmplot}
\end{figure*}

\section{Fisher matrix forecasts}

In order to quantify the power of peak number counts to constrain the HS model parameters, we carry on a Fisher matrix analysis considering as observed data the total number of peaks with ${\cal{S}} > {\cal{S}}_{th}$ in equally spaced redshift bins centred on $z$ and with width $\Delta z = 0.1$ over the range $(0.1, 1.4)$. As a first step, we need to define a likelihood function ${\cal{L}}$ to quantify how well a theoretical model matches the observed number counts. Should the errors be Gaussian, one can resort to the usual $\chi^2$ statistics and define $-2 \ln{{\cal{L}}} = \chi^2$. However, when dealing with number counts, one can assume Poisson errors\footnote{Note that, if there are uncertainties in assigning an object to a given bin, one should better resort to the {\it pull statistics} \citep{F02,C11}. If we assume that the peak redshift has been determined with sufficient accuracy, we can neglect this possibility and simply rely on the usual approach.} and then rely on the so called {\it C statistics} \citep{C79} to define\,:

\begin{equation}
-2 \ln{{\cal{L}}}({\bf p}) = - 2 \sum_{i = 1}^{{\cal{N}}_{bin}}{\kappa_i \ln{\lambda_i} - \lambda_i - \ln{\kappa_i !}}
\label{eq: deflike}
\end{equation}
where ${\bf p}$ is the set of model parameters, ${\cal{N}}_{bin}$ the number of redshift bins and, for notational clarity, we have defined $\lambda_i = {\cal{N}}_{pk}^{th}(z_i, {\bf p})$ and $\kappa_i = N_{pk}^{obs}(z_i)$ to denote the theoretical and observed number of peaks in the bin centred on $z_i$ (and with width $\Delta z = 0.1$). The Fisher matrix elements will then be given by the second derivatives of the logarithm of the likelihood with respect to the parameters of interest evaluated at the fiducial values. Starting from (\ref{eq: deflike}), one gets \citep{HHM01}\,:

\begin{equation}
F_{ij} = - \frac{\partial^2 \ln{{\cal{L}}}}{\partial p_i \partial p_j} =
\sum_{k = 1}^{{\cal{N}}_{bin}}{\frac{\partial \lambda_k}{\partial p_i} \frac{\partial \lambda_k}{\partial p_j} \frac{1}{\lambda_{k}^{fid}}}
\label{eq: fij}
\end{equation}
where $\lambda_{k}^{fid}$ is the expected number of peaks in the $k$\,-\,th bin for the fiducial model. The inverse of the Fisher matrix gives us the covariance matrix with its diagonal elements representing the lowest variance that one can achieve on the parameter $p_i$. It is worth noting that the Fisher matrix approximates confidence regions as a Gaussian ellipsoids, while the true confidence ones can have broad tails or significant curvature. However, \cite{HHM01} have checked that this is not the case for number counts by comparing with Monte Carlo analysis of simulated datasets. We expect that our estimated iso\,-\,likelihood contours provide a reliable approximation of the degeneracies in the parameters space.

Since here we are mainly interested in how peak number counts constrain $f(R)$ theories, we will not consider the full eight dimensional parameter space. On the contrary, we will set ${\bf p} = (\Omega_M, \varepsilon, n, \sigma_8)$, while $(h, w_0, w_a, n_{PS})$ are held fixed to their fiducial values. While this is not what one will actually do when dealing with real data, it is worth noting that here we are only considering peaks statistics as a constraint. Actually, the background quantities $(h, w_0, w_a)$ will be severely constrained by future SNeIa data, while $n_{PS}$ is determined by the fit to the CMBR spectrum. In a fully realistic approach, we should therefore both explore the full 8D parameter space and add the likelihood terms related to SNeIa and CMBR data. In order to better highlight the impact of peak statistics, we prefer to deal only with this kind of data and limit the parameter space. We therefore only consider $(\Omega_M, \varepsilon, n, \sigma_8)$ since they are intimately related to the mass function and the $f(R)$ model. The errors on these four quantities turn out to depend on the threshold $S/N$ used and are summarized in Table\,\ref{tab: fmres}, while Fig.\,\ref{fig: fmplot} shows the iso\,-\,likelihood contours in 2D spaces marginalizing over all the remaining parameters for the cases ${\cal{S}}_{th} = (4, 5)$.

\begin{table}
\caption{Fisher matrix forecasts for the accuracy on the parameters $(\Omega_M, \varepsilon, n, \sigma_8)$ from fitting the $N_{pk}(z, {\cal{S}} > {\cal{S}}_{th}$ data for different threshold $S/N$ ratio ${\cal{S}}_{th}$.}
\label{tab: fmres}
\begin{center}
\begin{tabular}{ccccc}
\hline
${\cal{S}}_{th}$ & $\Omega_M$ & $\varepsilon$ & $n$ & $\sigma_8$ \\
\hline \hline
%1 & 0.0279 & 0.0407 & 3.83 &  0.249 \\
2 & 0.0051 & 0.0452 & 1.12 &  0.068 \\
3 & 0.0007 & 0.2523 & 6.33 &  0.033 \\
4 & 0.0009 & 0.8189 & 18.8 &  0.042 \\
5 & 0.0020 & 1.0738 & 38.8 &  0.034 \\
\hline
\end{tabular}
\end{center}
\end{table}

It is interesting to note how the constraints scale with the threshold $S/N$. On the one hand, increasing ${\cal{S}}_{th}$ has the effect of reducing the overall number of peaks thus one expects a worsening of the constraints because of the poorer statistics. This is indeed the case for $(\varepsilon, n)$, but not for $(\Omega_M, \sigma_8)$. This is likely related to a second effect of a larger ${\cal{S}}_{th}$. As a consequence of the mass and redshift scaling of the $S/N$ ratio shown in Fig.\,\ref{fig: snratio}, varying ${\cal{S}}_{th}$ implicitly leads to probing a different MF regime. Since the dependence of the MF ${\cal{N}}(M_{vir}, z)$ on $(\Omega_M, \sigma_8)$ is different depending on the $(M_{vir}, z)$ region of the parameter space one is considering, the constraints will turn out to depend in a complicated way from which kind of peaks one is detecting.

Our main interest in the peaks statistics was motivated by constraining the $f(R)$ model parameters. Taken at face values, the constraints on $\varepsilon$ could look not so encouraging. We stress, however, that they are still competitive with those from an analysis based on background evolutionary probes as can be appreciated remembering that \cite{CCD11} obtained $\sigma(\varepsilon)/\varepsilon \sim 35\%$ from fitting SNeIa\,+\,GRBs\,+\,BAO\,+\,CMB distance priors vs the present forecasted $\sigma(\varepsilon)/\varepsilon \sim 16\%$ value (for ${\cal{S}}_{th} = 5$) using weak lensing peaks counts only.

As it is apparent from both the numbers and the plots, the only parameter which remains fully unconstrained is the slope $n$ of the HS model. We could naively anticipate this result by simply looking at the top right panel of Fig.\,\ref{fig: npkpar} where one can see that $N_{pk}(z)$ changes by less than $0.1\%$ up to $z \simeq 1$ so that this parameter has a negligible impact on the observed peak number. This conclusion can be qualitatively explained by looking at the effective gravitational constant ${\cal{G}}_{eff}$ in Eq.(\ref{eq: geffdef}). For the $\varepsilon$ values typical of viable HS models, the dependence of ${\cal{G}}_{eff}$ on $n$ almost cancels out from the ratio so that the power spectrum $P_{\delta}(k, z)$ is almost independent of $n$ with the residual sensitivity being smoothed out from the integration needed to get the variance $\sigma^2(M_{vir})$. It is therefore not surprising that the Fisher matrix analysis finally leads to a large uncertainty on $n$. We, nevertheless, caution the reader that this is partly a consequence of how we have set the fiducial model. Indeed, as already hinted at in Sect.\,2.3, for $(n, \varepsilon) = (1.5, -6.0)$, the $f(R)$ Lagrangian is quite similar to the $\Lambda$CDM one with the term depending on $n$ giving an almost negligible contribution. Investigating how the Fisher matrix forecasts change with the fiducial model is, however, outside our aims here.

The stronger dependence of $N_{pk}(z)$ on $\varepsilon$ which can be read from the top left panel of Fig.\,\ref{fig: npkpar}, on the contrary, helps us to understand why peaks statistics is so efficient in constraining this parameter. Again, we can look at ${\cal{G}}_{eff}$ to qualitatively explain why the results so strongly depend on $\varepsilon$. First, we note that, for the values of interest, we get ${\cal{G}}_{eff} \simeq G/(1 - 10^{\varepsilon})$ within a good approximation. Second, the interpolation between $\sigma_{FoG}$ and $\sigma_{GR}$ in Eq.(\ref{eq: sigmalh}) is mediated by a mass scale which depends on $\varepsilon$. Small variation in this quantity can therefore significantly change the number of peaks thus leading to stronger constraints.

It is worth stressing that the constraints in Table\,\ref{tab: fmres} are actually quite optimistic since they are based on the assumption that one is able to partition the detected peaks in redshift bins. First, we note that, should the peak be due to the LSS, it will not have any actual counterpart in optical images so that the notion of redshift can not be applied. We are therefore forced to consider only the results for the case ${\cal{S}}_{th} = 5$ (or larger). Even limiting our attention to these high ${\cal{S}}$ peaks only,  there is still the problem of how to divide them in redshift bins. Ideally, one should rely on optical clusters finders \citep{Post96,Koe07,Milk10,Bella11} to match the peaks catalog with optical detections and then infer the redshift. {\bf To this end, it is worth noting that most of the optical finders techniques rely on the same kind of data as those available for Euclid so that the match with weak lensing detections could be immediate. Moreover, since the cluster redshift estimate is based on more galaxies at once supplemented by geometrical constraints, the precision on $z$ will be similar (if not higher) than that of the photometric redshift of a single galaxy.} In a first approximation, we can therefore assume that the optical finder provides a probability distribution function for $z$ which can be modelled as a Gaussian with negligible bias and variance $\sigma_z = \sigma_0 (1 + z)$. In order a peak with $z_i \le z \le z_{i + 1}$ not to be incorrectly assigned to a different redshift bin, we can roughly ask that the $3\sigma$ uncertainty on $z$ is smaller than the bin width. For our assumed value $\Delta z = 0.1$, this translates in $\sigma_z \le 0.03 (1 + z)$. Such a precision could be likely achieved if $z$ is spectroscopically measured ($\sigma_z < 0.001 (1 + z)$ according to the Euclid red book), but could be too demanding if one relies on photometric redshift methods ($\sigma_z < 0.05 (1 + z)$ for Euclid). In this second case, one should repeat the above Fisher matrix analysis adding a non Poissonian uncertainty on $N_{pk}(z, {\cal{S}} > {\cal{S}}_{th})$ and resorting to the pull statistics \citep{F02,C11}. Whatever is the method adopted for inferring the peak redshift, it is likely that its precision will also depend on the peak $S/N$ ratio so that the net effect could be included in our analysis by convolving the theoretically computed $N_{pk}(z)$ with an empirically determined selection function. Investigating this issue is outside our aims here so that we only warn the reader that the numbers in Table\,\ref{tab: fmres} should be taken as (likely optimistic) upper limits.

\begin{figure}
\centering
\includegraphics[width=7.5cm]{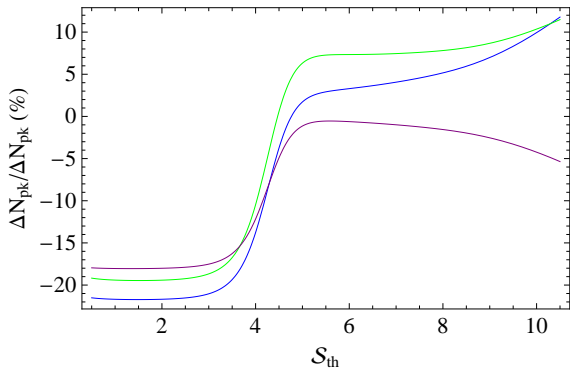}
\caption{Percentate deviation of the total number of peaks from the fiducial model for $\varepsilon = -4$ (green), $-5$ (blue), $-7$ (purple).}
\label{fig: ntotepslog}
\end{figure}

As an alternative, one could rely on the total number of peaks as only observational constraint. As shown in Fig.\,\ref{fig: ntotepslog}, this quantity has a significant dependence on $\varepsilon$ so that it can be a valuable help to narrow down the range for this parameter. However, a single quantity is unable to put constraints in a 4D space so that we have not carried out a Fisher matrix analysis for this case. We stress, however, that meaningful constraints could be obtained adding background probes such as, e.g., SNeIa and GRBs. As an example, we can remember that in \cite{CCD11}, we found $\-7.42 \le \varepsilon \le -3.48$ at the $68\%$ CL. Should one determine $N_{pk}({\cal{S}} > {\cal{S}}_{th})$ to be consistent with the fiducial $N_{pk}$ value within $10\%$ for ${\cal{S}}_{th} = 2$, one could safely exclude the three models in Fig.\,\ref{fig: ntotepslog} thus greatly narrowing the confidence range. A joint analysis is, however, needed to investigate the role of degeneracies with other HS model parameters.

\section{Conclusions}

The unexpected discovery of the cosmic acceleration promptly raised numerous papers proposing different candidates to drive the accelerated expansion. It soon became clear that one can also go beyond General Relativity to modified gravity theories. In particular, $f(R)$ theories were shown to be able to provide exactly the same background expansion of dark energy models so that the issue nowadays is, more than checking their validity, discriminating between dark energy and modified gravity. Here, we have shown that this is in principle possible by relying on the peak statistics, i.e. the number of peaks in the weak lensing maps constructed using the mass aperture statistics. In order to quantify this possibility, we have carried on a Fisher matrix analysis to estimate the accuracy on the model parameters which an Euclid\,-\,like survey can achieve using peak number counts as the only observational constraint.

Although the results are quite encouraging, it is worth asking whether they can be improved and refined. First, we note that the theoretical quantity we have actually relied on is the mass function. One can repeat our analysis by using different tracers of the mass function itself such as, e.g., X\,-\,ray catalogues. Number counts of X\,-\,ray selected clusters have indeed been widely investigated as a possible tool for precision cosmology \citep{HMH01,Mantz08,BPL10,Pierre11} so that one can naively believe that a joint analysis of the mass function of X\,-\,ray and weak lensing selected clusters can improve the constraints on the $f(R)$ model parameters. Actually, validating such a prediction is not straightforward. On the one hand, since both kinds of observations probe the same quantity it is indeed possible that degeneracies in the model parameter space are not broken so that no significative improvement is achieved. On the other hand, while lensing probes the full matter distribution notwithstanding its dynamical state, X\,-\,ray number counts are subject to the uncertainties related to the cluster mass determination which typically relies on scaling relations which have been determined in a General Relativity framework. Although great care has to be used to take into account this problem, we nevertheless believe that such a task is worth being carried out in order to complement the peak statistics.

As a second issue, we remind the reader that we have only investigated a subset of the full 8D parameter space. Allowing for a larger number of parameters to be varied impacts the constraints by introducing further degeneracies and hence widening the confidence ranges. In order to compensate for this degradation of the constraining power of the method, one can complement peak statistics with different dataset. From this point of view, SNeIa and CMBR are ideal tools. Indeed, the SNeIa Hubble diagram is a reliable tracer of the expansion rate over approximately the same redshift range covered by the peak number counts; this offers the possibility to severely constrain those parameters most related to the distance vs redshift relation such as $(h, w_0, w_a)$. Similarly, the CMBR anisotropy spectrum strongly depends on $(n, \sigma_8)$ so that fitting this dataset allows to set these parameters and hence break the $(\Omega_M, \sigma_8)$ and $(\sigma_8, \varepsilon)$ degeneracies shown in Fig.\,\ref{fig: fmplot}.

As a final comment, we would like to stress that cosmic shear tomography can be efficiently added to peak number counts since it is has been shown to be particularly efficient at constraining the HS parameters \citep{CDC11}. Considering that both the shear power spectrum and the peak number counts rely on the same underlying phenomenon (the lensing distortion of the images of background galaxies from an intervening mass distribution) and both will be measured from the Euclid mission, we end up with the intriguing possibility to use a single mission and a single probe to give a definitive answer to the up to now unsolved dark energy vs modified gravity controversy.

\section*{Acknowledgements}

VFC warmly thanks M. Maturi for his prompt answers on the optimal filter. We also kindly acknowledge the referee for his/her insightful comments which have helped us to improve the paper. VFC is funded by the Italian Space Agency (ASI) through contract Euclid\,-\,IC (I/031/10/0). VFC and RS acknowledge financial contribution from the agreement ASI/INAF I/023/12/0. SC acknowledges support from FCT\,-\,Portugal under grant PTDC/FIS/100170/2008. SC's work is funded by FCT\,-\,Portugal under Post\,-\,Doctoral Grant SFRH/BPD/80274/2011. AD acknowledges partial support from the INFN grant PD51  and the PRIN-MIUR-2008 grant \verb"2008NR3EBK_003" ``Matter-antimatter asymmetry, dark matter and dark energy in the LHC era''.


\begin{thebibliography}{99}

\bibitem[\protect\citeauthoryear{Amanullah et al. }{2010}]{Union2}
Amanullah, R., Lidman, C., Rubin, D., Aldering, G., Astier, P., et al. 2010, ApJ, 716, 712

\bibitem[\protect\citeauthoryear{Bartelmann }{1996}]{B96}
Bartelmann, M. 1996, A\&A, 313, 697

\bibitem[\protect\citeauthoryear{Bartelmann et al. }{2002}]{BPB02}
Bartelmann, M., Perrotta, F., Baccigalupi, C. 2002, A\&A, 396, 21

\bibitem[\protect\citeauthoryear{Basilakos et al. }{2010}]{BPL10}
Basilakos, S., Plionis, M., Lima, J.A.S. 2010, Phys. Rev. D, 82, 083517

\bibitem[\protect\citeauthoryear{Bryan \& Norman }{1998}]{BN98}
Bryan, G.L., Norman, M.L. 1998, ApJ, 495, 80

\bibitem[\protect\citeauthoryear{Bellagamba et al. }{2011}]{Bella11}
Bellagamba, F., Maturi, M., Hamana, T., Meneghetti, M., Miyazaki, S., Moscardini, L. 2011, MNRAS, 413, 1145

\bibitem[\protect\citeauthoryear{Buote et al. }{2007}]{B07}
Buote, D.A., Gastaldello, F., Humphrey, P.J., Zappacosta, L., Bullock, J.S., Brighenti, F., Mathews, W.G. 2007, ApJ, 664, 123

\bibitem[\protect\citeauthoryear{Bullock et al. }{2001}]{B01}
Bullock, J.S., Kolatt, T.S., Sigad, Y., Somerville, R.S., Kravtsov, A.V., Klypin, A.A., Primack, J.R., Dekel, A. 2001, MNRAS, 321, 559

\bibitem[\protect\citeauthoryear{Cash }{1979}]{C79}
Cash, W. 1979, ApJ, 228, 939

\bibitem[\protect\citeauthoryear{Camera et al. }{2011}]{CDC11}
Camera, S., Diaferio, A., Cardone, V.F. 2011, JCAP, 07, 016

\bibitem[\protect\citeauthoryear{Campanelli et al. }{2011}]{C11}
Campanelli, L., Fogli, G.L., Kahniashvili, T., Marrone, A., Ratra, B. 2011, preprint arXiv\,:1110.2310

\bibitem[\protect\citeauthoryear{Capozziello et al. }{2005}]{CCT05}
Capozziello, S., Cardone, V.F., Troisi, A. 2005, Phys. Rev. D, 71, 043503

\bibitem[\protect\citeauthoryear{Capozziello et al. }{2008}]{CCS08}
Capozziello, S., Cardone, V.F., Salzano, V. 2008, Phys. Rev. D, 78, 063504

\bibitem[\protect\citeauthoryear{Capozziello \& de Laurentis }{2011}]{CdL11}
Capozziello, S., de Laurentis, M. 2011, Phys. Rept., 509, 167

\bibitem[\protect\citeauthoryear{Cardone \& Capozziello }{2011}]{CC11}
Cardone, V.F., Capozziello, S. 2011, MNRAS, 414, 1301

\bibitem[\protect\citeauthoryear{Cardone et al. }{2012}]{CCD11}
Cardone, V.F., Camera, S., Diaferio, A. 2012, JCAP, 02, 030

\bibitem[\protect\citeauthoryear{Chevallier \& Polarski }{2001}]{CP01}
Chevallier, M., Polarski, D. 2001, Int. J. Mod. Phys. D, 10, 213

\bibitem[\protect\citeauthoryear{Clowe et al. }{1998}]{Clowe98}
Clowe, D., Luppino, G.A., Kaiser, N., Henry, J.P., Gioia, I.M. 1998, ApJ, 479, L61

\bibitem[\protect\citeauthoryear{Dahle }{2006}]{D06}
Dahle, H. 2006, ApJ, 653, 954

\bibitem[\protect\citeauthoryear{de Felice \& Tsujikawa }{2010}]{dFT10}
de Felice, A., Tsujikawa, S. 2010, Liv. Rev. Rel., 13, 3

\bibitem[\protect\citeauthoryear{Dev et al. }{2008}]{dev}
Dev, A., Jain, B., Jhingan, S., Nojiri, S., Sami, M., Thongkool, I. 2008, Phys. Rev. D, 78, 083515

\bibitem[\protect\citeauthoryear{Dietrich \& Hartlap }{2010}]{DH10}
Dietrich, J.P., Hartlap, J. 2010, MNRAS, 402, 1049

\bibitem[\protect\citeauthoryear{Eisenstein \& Hu }{1998}]{EH98}
Eisenstein, D.J., Hu, W. 1998, ApJ, 496, 605

\bibitem[\protect\citeauthoryear{Faulkner et al. }{2007}]{F07}
Faulkner, T., Tegmark, M., Bunn, E.F., Mao, Y. 2007, Phys. Rev. D, 76, 0603505

\bibitem[\protect\citeauthoryear{Fogli et al. }{2002}]{F02}
Fogli, G.L., Lisi, E., Marrone, A., Montanino D. 2002, Phys. Rev. D, 66, 053010

\bibitem[\protect\citeauthoryear{Gavazzi \& Soucail }{2007}]{GS07}
Gavazzi, R., Soucail, G. 2007, A\&A, 462, 459

\bibitem[\protect\citeauthoryear{Geller et al. }{2010}]{G10}
Geller, M.J., Kurtz, M.J., Dell'\,Antonio, I.P., Ramella, M., Fabricant, D.G. 2010, ApJ, 709, 832

\bibitem[\protect\citeauthoryear{Hetterscheidt et al. }{2005}]{H05}
Hetterscheidt, M., Erben, T., Schneider, P., Maoli, R., van Waerbecke, L., Mellier, Y. 2005, A\&A, 442, 43

\bibitem[\protect\citeauthoryear{Hoekstra }{2001}]{H01}
Hoekstra, H. 2001, A\&A, 370, 743

\bibitem[\protect\citeauthoryear{Holder et al. }{2001a}]{HHM01}
Holder, G., Haiman, Z., Mohr, J.J. 2001a, ApJ, 560, L111

\bibitem[\protect\citeauthoryear{Holder et al. }{2001b}]{HMH01}
Holder, G., Mohr, J.J., Haiman, Z. 2001b, ApJ, 553, 545

\bibitem[\protect\citeauthoryear{Hu \& Sawicki }{2007}]{HS07}
Hu, W., Sawicki, I. 2007, Phys. Rev. D, 76, 064004

\bibitem[\protect\citeauthoryear{Huang et al. }{2011}]{Zhuoyi11}
Huang, Z., Radovich, M., Grado, A., Puddu, E., Romano, A., Limatola, L., Fu, L. 2011, A\&A, 529, 39

\bibitem[\protect\citeauthoryear{Kaiser \& Squires }{1993}]{KS93}
Kaiser, N., Squires, G. 1993, ApJ, 404, 441

\bibitem[\protect\citeauthoryear{Kaiser et al. }{1995}]{KSB95}
Kaiser, N., Squires, G., Broadhurst, T. 1995, ApJ, 449, 460

\bibitem[\protect\citeauthoryear{Khoury \& Weltman }{2004}]{KW04}
Khoury, J., Weltman, A. 2004, Phys. Rev. D, 69, 044026

\bibitem[\protect\citeauthoryear{Koester et al. }{2007}]{Koe07}
Koester, B.P., McKay, T.A., Annis, J., Wechsler, R.H., Evrard, A., Bleem, L., Becker, M., Johnston, D. 2007, ApJ, 660, 239

\bibitem[\protect\citeauthoryear{Komatsu et al. }{2011}]{WMAP7}
Komatsu, E., Smith, K.M., Dunkley, J., Bennett, C.L., Gold, B. et al. 2011, ApJS, 192, 18

\bibitem[\protect\citeauthoryear{Kratochvil et al. }{2010}]{K10}
Kratochvil, J.M., Haiman, Z., May, M. 2010, Phys. Rev. D, 81, 043519

\bibitem[\protect\citeauthoryear{Kurtz et al. }{2012}]{K12}
Kurtz, M.J., Geller, M.J., Utsumi, Y., Miyazaki, S., Dell'\,Antonio, I.P., Fabricant, D.G. 2012, ApJ, 750, 168

\bibitem[\protect\citeauthoryear{Laurejis et al. }{2011}]{RB}
Laureijs, R., Amiaux, J., Arduini, S., Augu\`{e}res, J.L., Brinchmann, J., et al. 2011, preprint arXiv\,:1110.3193

\bibitem[\protect\citeauthoryear{Li \& Hu }{2011}]{LH11}
Li, Y., Hu, W. 2011, Phys. Rev. D, 84, 084033

\bibitem[\protect\citeauthoryear{Linder }{2003}]{L03}
Linder, E.V. 2003, Phys. Rev. Lett., 90, 091301

\bibitem[\protect\citeauthoryear{Mantz et al. }{2008}]{Mantz08}
Mantz, A., Allen, S.W., Ebeling, H., Rapetti, D. 2008, MNRAS, 387, 1179

\bibitem[\protect\citeauthoryear{Marian et al. }{2009}]{M09}
Marian, L., Smith, R.E., Bernstein, G.M. 2009, ApJ, 698, L33

\bibitem[\protect\citeauthoryear{Marian et al. }{2011}]{M11}
Marian, L., Hilbert, S., Smith, R.E., Schneider, P., Desjacques, V. 2011, ApJ, 728, L13

\bibitem[\protect\citeauthoryear{Maturi et al. }{2005}]{M05}
Maturi, M., Meneghetti, M., Bartelmann, M., Dolag, K., Moscardini, L. 2005, A\&A, 442, 851

\bibitem[\protect\citeauthoryear{Maturi et al. }{2010}]{M10}
Maturi, M., Angrick, C., Pace, F., Bartelmann, M. 2010, A\&A, 519, A23

\bibitem[\protect\citeauthoryear{Maturi et al. }{2011}]{MFM11}
Maturi, M., Fedeli, C., Moscardini, L. 2011, 416, 2527

\bibitem[\protect\citeauthoryear{Milkeraitis et al. }{2010}]{Milk10}
Milkeraitis, M., van Waerbeke, L., Heymans, C., Hildebrandt, H., Dietrich, J.P., Erben, T. 2010, MNRAS, 406, 673

\bibitem[\protect\citeauthoryear{Mota \& Barrow }{2004}]{MB04}
Mota, D.F., Barrow, J.D. 2004, Phys. Lett. B, 581, 141

\bibitem[\protect\citeauthoryear{Multam\"aki \& Vilja }{2006}]{MV06}
Multam\"aki, T., Vilja, I. 2006, Phys. Rev. D, 73, 024018

\bibitem[\protect\citeauthoryear{Mu\~noz\,-\,Cuartas et al. }{2011}]{MC11}
Mu\~noz\,-\,Cuartas, J.C., Macci\`o, A.V., Gottl\"ober, S., Dutton, A. 2011, MNRAS, 411, 584

\bibitem[\protect\citeauthoryear{Navarro et al. }{1997}]{NFW97}
Navarro, J.F., Frenk, C.S., White, S.D.M. 1997, ApJ, 490, 493

\bibitem[\protect\citeauthoryear{Percival et al. }{2010}]{P10}
Percival, W.J., Reid, B.A., Eisenstein, D.J., Bahcall, N.A., Budavari, T., et al. 2010, MNRAS, 401, 2148

\bibitem[\protect\citeauthoryear{Pierre et al. }{2011}]{Pierre11}
Pierre, M., Pacaud, F., Juin, J.B., Melin, J.B., Valageas, P., Clerc, N., Corasaniti, P.S. 2011, MNRAS, 414, 1732

\bibitem[\protect\citeauthoryear{Postman et al. }{1996}]{Post96}
Postman, M., Lubin, L.M., Gunn, J.E., Oke, J.B., Hoessel, J.G., Schneider, D.P., Christensen, J.A. 1996, ApJ, 111, 615

\bibitem[\protect\citeauthoryear{Radovich et al. }{2008}]{Mario2008}
Radovich, M., Puddu, E., Romano, A., Grado, A., Getman, F. 2008, A\&A, 487, 55

\bibitem[\protect\citeauthoryear{Romano et al. }{2010}]{Anna10}
Romano, A., Fu, L., Giordano, F., Maoli, R., Martini, P., et al. 2010, A\&A, 514, 88

\bibitem[\protect\citeauthoryear{Schirmer et al. }{2004}]{S04}
Schirmer, M., Erben, T., Schneider, P., Wolf, C., Mesenheimer, K. 2004, A\&A, 420, 75

\bibitem[\protect\citeauthoryear{Schirmer et al. }{2007}]{Sch07}
Schirmer, M., Erben, T., Hetterscheidt, M., Schneider, P. 2007, A\&A, 462, 875

\bibitem[\protect\citeauthoryear{Schmidt et al. }{2009}]{S09}
Schmidt, F., Lima, M., Oyaizu, H., Hu, W. 2009, Phys. Rev. D, 79, 083518

\bibitem[\protect\citeauthoryear{Schneider }{1996}]{S96}
Schneider, P. 1996, MNRAS, 283, 837

\bibitem[\protect\citeauthoryear{Schneider et al. }{1998}]{S98}
Schneider, P., van Waerbecke, L., Jain, B., Kruse, G. 1998, MNRAS, 296, 873

\bibitem[\protect\citeauthoryear{Sheth \& Tormen }{1999}]{ST99}
Sheth, R., Tormen, G. 1999, MNRAS, 308, 119

\bibitem[\protect\citeauthoryear{Soucail et al. }{1987}]{S87}
Soucail, G., Fort, B., Mellier, Y., Picat, J.P. 1987, A\&A, 184, L14

\bibitem[\protect\citeauthoryear{Tsujikawa }{2007}]{T07}
Tsujikawa, S. 2007, Phys. Rev. D, 76, 023514

\bibitem[\protect\citeauthoryear{Webster }{1985}]{W85}
Webster, R.L. 1985, MNRAS, 213, 871

\bibitem[\protect\citeauthoryear{Weinberg et al. }{2012}]{W12}
Weinberg, D., Mortonson, M.J., Eisenstein, D.J., Hirata, C., Riess, A., Rozo, E. 2012, preprint arXiv\,:1201.2434

\bibitem[\protect\citeauthoryear{Wright \& Brainerd }{2000}]{WB00}
Wright, C.O., Brainerd, T.G. 2000, ApJ, 534, 34

\end{thebibliography}
\end{document}